\documentclass[aps,prd,onecolumn,eqsecnum,amsmath,nofootinbib,preprintnumbers]{revtex4}%

\usepackage{color,graphicx,float,subfigure,xcolor}
\usepackage{amsfonts,amssymb,theorem,mathrsfs,times}
\usepackage{bm}
\usepackage{mathtools}
\usepackage{amsfonts,amssymb,theorem,mathrsfs}
\usepackage{dsfont}
\usepackage{setspace}
\usepackage{amsmath}  
\usepackage{xcolor}
\usepackage[colorlinks,linkcolor=blue,anchorcolor=blue,citecolor=blue]{hyperref}   
\usepackage{ulem}   
\usepackage[english]{babel}

{\theorembodyfont{\upshape}
	}
{\theorembodyfont{\upshape}
	}
{\theorembodyfont{\upshape}
	}
{\theorembodyfont{\upshape}
	}
{\theorembodyfont{\upshape}
	}
{\theorembodyfont{\upshape}
	}
\addtocounter{MaxMatrixCols}{10}
\newcommand{\dalm}{\kern1pt\vbox{\hrule height 0.9pt\hbox{\vrule width
			0.9pt\hskip 2.5pt\vbox{\vskip 5.5pt}\hskip 3pt\vrule width
			0.3pt}\hrule height 0.3pt}\kern1pt}

\begin{document}
\thispagestyle{empty}
\preprint{\hfill {\small {ICTS-USTC/PCFT-24-03}}}
\title{The quasinormal modes, pseudospectrum and time evolution of Proca fields in quantum Oppenheimer-Snyder–de Sitter spacetime}
%
\author{ Shu Luo$^{a}$\footnote{e-mail
			address: ls040629@mail.ustc.edu.cn}}

\affiliation{${}^a$School of  Physics ,
 University of Science and Technology of China, Hefei, Anhui 230026,
	China}

\date{\today}
	
\begin{abstract}
In this study, we investigate the quasinormal modes, pseudospectrum and time evolution of a massive vector field around a quantum corrected black hole in de-Sitter spacetime. We start by parameterization and using orthonormal tetrads to get the effective potential. Methodologically we use the hyperboloidal framework together with discretizing the non-selfadjoint operator through Chebyshev-Gauss-Labatto grid to attain the QNMs. We explore the parametric instability of QNMs caused by quantum correction, cosmological constant and Proca mass, and these three factors show very different influences on the QNMs' migration flow. On the other hand, we discuss the instability of QNMs with arbitrary-shape perturbation and the effectiveness of numerical results through pseudospectrum. We use high frequency approximation to attain the expression of the time domain Green function and clarify the origin of two different stages in time evolution. Through numerical methods we confirm that no power-law late time tail is expected, and the possible impact on time evolution caused by quantum correction is discussed. 
\end{abstract}

\maketitle

\section{Introduction}
Recently, with the detection of gravitational waves(GW) from 
binary compact stars~\cite{LIGOScientific:2016aoc}, together with the Event Horizon Telescope (EHT) revealing the optical characteristics of M87 center supermassive black hole~\cite{Akiyama_2019}, black holes, especially astronomical realistic black holes, can serve as a lab to test various theoretical predictions through Multi Messenger Astronomy, including quantum gravity and other modified gravity theories . These modifications mainly consists of two types: adding new elements to the right hand of the Einstein equation, i.e., supposing new source of energy-momentum tensor to tangle with gravitation, including dark matter and dark energy~\cite{Salucci_2000,Burkert_1995}; modifying the left side of the equation by the new gravitational degrees of freedom in addition to the metric tensor. In many cases, the modification of GR can be described by a scalar-tensor theory of gravitation~\cite{CLIFTON20121,Berti_2015}. Apart from the extensively-investigated scalar-tensor theory, there is another class of generalized vector-tensor theories, Einstein-Proca theories, based on the generalized Proca equation and higher order coupling of gravitation and Proca fields~\cite{Allys_2016,BELTRANJIMENEZ2016405}, and there are investigations on relational spherically symmetric solutions~\cite{PhysRevD.94.084039}, superradiant instabilities which offer restrict on the upper bounds of photons~\cite{PhysRevD.87.043513,PhysRevLett.109.131102}, vector field quasinormal modes~\cite{PhysRevLett.120.231103}  and optical characteristics~\cite{PhysRevD.99.024052}. 

Among these efforts, one important problem to be solved by new theories is spacetime singularities, characterized by infinite curvature or density, which has been a subject of great interest and curiosity in the fields of gravitation theory and relativistic astrophysics. The existence of singularities is
unavoidable in generic gravitational collapses, which has been proven by Hawking and Penrose in the most general case. The presence of singularities poses profound challenges to our
understanding of the universe within the context of classical
general relativity, however, a probable candidate for quantum gravity theory, loop quantum gravity (LQG), provides a different view of this problem. It claims that through quantum geometry the singularity no matter in a black hole or the beginning of our universe can be solved~\cite{PhysRevD.107.064011,STACHOWIAK2007209}. 

Recently, Lewandowski\,\textit{et al.}\,have proposed a new quantum black hole model (qOS) within the framework of Loop Quantum Gravity (LQG) theory~\cite{PhysRevLett.130.101501}. The exterior spacetime of this model is a suitably deformed Schwarzschild black hole. This quantum corrected black hole exhibits the same asymptotic behavior as the Schwarzschild black hole and it is stable against test scalar and vector fields by the analysis of QNMs~\cite{PhysRevD.108.104004}. Through the similar method, the solution describing a quantum black hole with a positive cosmological constant (qOS-dS) is obtained in~\cite{PhysRevD.109.064012}. 

Although having a lot of similarities to the common case,  such a solution still has several observable properties that are different from those of the Schwarzschild-dS black hole, including its shadow and photon sphere, and by adding the cosmological constant, the authors in~\cite{PhysRevD.109.064012} focus on checking the SCC hypothesis in this spacetime. They put forward that SCC hypothesis will be destroyed as the black hole approaches the near-extremal limit. Additionally, in~\cite{PhysRevD.109.064012} some aspects about the quasinormal modes (QNMs) in the spacetime are also studied, but with few discussions on the influence of the perturbations of effective potential and the (in)stability of the QNMs, or  quantitative description of the instability through pseudospectrum method.

The studies of quasinormal modes (QNMs) mainly focus on characteristic behaviors of the modes under different types of spacetime. Starting from the 1970s, a lot of different spin fields' QNMs in a lot of kinds of D-dimensional spacetime have been deeply studied~\cite{RevModPhys.83.793}. Another center of investigation is the pseudospectrum and instability of the QNM, starting from the pioneering work of Nollert~\cite{PhysRevD.53.4397}, which has shown that the QNM overtones are strongly unstable with their instabilities increasing with their damping~\cite{PhysRevD.101.104009,PhysRevD.103.024019}. Recent works focused on this topic include \cite{Konoplya_2022,konoplya2022overtonesprobeeventhorizon}. A very tiny parameter modification might sharply influence the behavior of QNMs, especially for the case of the comparison between a strictly massless field and very small but nonzero massive fields. It has been shown that massive fields own some modes that almost do not decay, corresponding to imaginary part very close to zero, which is called ``quasi-resonances"~\cite{RevModPhys.83.793,KONOPLYA2005377}. This behavior very likely happens when the amplitude near the event horizon is even smaller than far from the black hole, which means there is no leak of energy from the system and there exist some modes of standing waves. Theoretical analyses also support purely real modes when there is a massive field~\cite{KONOPLYA2005377}.

We mainly investigate the QNMs, their instability and evolution in the time domain of the quantum corrected black hole (qOS-dS) in this paper, and we choose the Proca field as the test field and explore their axial perturbation.The quantization of Proca field has been done in~\cite{Helesfai_2007}, in which the author presented a new method to solve the problem caused by the secondary constraints characteristics of Proca field in Loop Quantum Gravity (LQG). Meanwhile, in~\cite{Helesfai_2008}, the author made clear the relation between the well-known symmetric breaking and the massive vector field in the framework of LQG. The Hamiltonian constructed for Proca field in LQG was found to have good characteristics, with no UV divergence or unnatural treatment of renormalisation but by merely treating gravity dynamical.For this reason we choose Proca field to serve as the test field of a modified metric under LQG. This means we will combine the three important study center mentioned in the above paragraphs. And we will summarize the influence of these factors on QNMs by three parameters: the relative ratio of Cauchy horizon radius to outer horizon radius $q$, the ratio of outer horizon radius to cosmological horizon radius $p$ and Proca mass $m$. The first reason of choosing this kind of field is to confirm the origin of the quasi-resonance phenomenon~\cite{RevModPhys.83.793,KONOPLYA2005377} in a non asymptotic flat spacetime, and it will be shown that particularly for vector field the perturbation caused by changing Proca mass $m$ can be analogized to continuous migration of the angular momentum number $l$, but $m$ no longer plays the important role of deciding the asymptotic behavior, so no quasi-resonance is found. More practically, the detection on the Ringdown stage of the time domain (which has been known to closely dependant on QNMs) to ensure the upper limit of the quantum correction may be disturbed by the possible mass term of the photons, so it's important to clarify the influence caused by mass term.

The first step of our study is the parameterization of the whole metric using the parameters $p$ and $q$ in the last paragraph, which can simplify the following investigations on the perturbations caused by quantum correction, excluding nonphysical range of parameters that do not accord to the premise discussed. As we need to construct a hyperboloidal coordinate to absorb the asymptotic boundary condition into non-divergent coordinate condition, compared to previous investigation~\cite{PhysRevD.109.064012} this parametric form seems to be more convenient. Then the governing equation of the axial perturbation of the Proca field in qOS-dS spacetime is obtained, together with using hyperboloidal framework to attain the QNMs. Moreover, in this process a possibility of losing the effectiveness of this method under the minimal gauge is discussed, which is of important physical meaning and offers a useful warning in later exploration.

The instability of quasinormal modes can be quantified from two perspectives. On the one hand, the solution as a  correction to the Schwarzschild is used to discuss spectrum (in)stability of Schwarzschild black hole. In this way we will clarify the total tendency of the frequency domain when some parameters migrate. And some important characteristics under quantum correction is revealed. On the other hand, we will use the pseudospectrum method to study the relative stability of the QNMs of different overtones under arbitrary perturbation of a fixed amplitude, which has never been done before for this solution.

After all these discussions are finished, the evolution in the time domain is also discussed in detail. We mainly focus on the different stages of the evolution. The lasting time of the first stage, Precursor, is influenced by parameters. The second stage,  Ringdown, also show some important signals affected by quantum correction. About the third stage, as mentioned in~\cite{PhysRevD.106.064058}, the late-time tails can reflect some essential properties of black
holes, such as the no-hair theorem and the instability of
Cauchy horizons, and here under a non-zero cosmological constant it strictly obeys exponentially decaying rule. Also, a detail about the distribution of the energy on the spatial domain is incidentally mentioned afterwards.

The paper is organized as follows. In the next section we finish the parameterization of the qOS-dS spacetime. Then in Sec.\ref{sec:2}, we present the equation of motion of Proca field in qOS-dS spacetime. In Sec.\ref{sec:3} we construct the hyperboloidal coordinate to calculate QNMs numerically. In Sec.\ref{sec:41} we discuss the instability of the QNMs, first by parametric migration in Sec.\ref{sec:10} and then by pseudospectrum in Sec.\ref{sec:42}. We then attain the high-frequency approximation of the Green function and numerically calculate the evolution in time domain in Sec.\ref{sec:4}. Sec.\ref{sec:5} is the conclusions and discussion.

\section{parameterization of The qOS-dS solution}\label{sec:1}

We first briefly discuss the physical image of a qOS-dS model. Through the GR, it has been revealed that a collapsing star can be described by the Oppenheimer-Snyder model, which predicts that a singularity in the final evolutionary stage of the star is unavoidable. However, the loop quantum theory (LQG) predicts a different picture of the collapsing and bouncing matters in a certain dust ball of a time-changing radius never smaller than a certain $r_{b}$. According to the Ashtekar-Pawlowski-Singh (APS) model, the inner region of a such dust ball can be described by an non-globally static metric in the same form of the Robertson-Walker metric with $k=0$, while the exterior spacetime is a pseudo-static one, which has a form close to Schiwarzschild spacetime. The only difference lies in the dynamical equation ruling the evolution of the scale parameter, or, equivalently, the radius of the dust ball.

 And the mere difference between the qOS-dS and qOS model is the non-vanishing cosmological constant, which contributes to the dynamical evolution of the scale factor $a(\tau)$, and the expression of the exterior spacetime is also modified. Specifically speaking, the metric has the form of
 \begin{eqnarray}
     \mathrm{d}s_{\text{in}}^2=-\mathrm{d}\tau^2+a(\tau)^2[\mathrm{d}\tilde{r}^2+\tilde{r}^2(\mathrm{d}\theta^2+\mathrm{sin}^2\theta \mathrm{d}\phi^2)]\,,
 \end{eqnarray}
 inside the dust bull surface, in which the scale factor is governed by  (we have set gravity constant $G=1$)
\begin{eqnarray}
    H=\Big{(}\frac{\dot{a}}{a}\Big{)}^2=\frac{8\pi}{3}\rho(1-\frac{\rho}{\rho_{c}})+\frac{\Lambda}{3}\,,\qquad \rho=\frac{M}{\frac{4\pi}{3}a^3\tilde{r}_{0}^3}\,,
\end{eqnarray}
in which $\rho_{c}=\sqrt{3}/32\pi^2\gamma^3l_{p}^2$, $l_{p}=\sqrt{\hbar}$ is the Plank length, $\gamma$ being the Immirzi parameter, $\Lambda$ being the cosmological constant and $M$ standing for the mass of the dust ball. The coordinate $\tilde{r}$ only makes sense in the area $0<\tilde{r}<\tilde{r}_{0}$.

\begin{figure}[htbp]
	\centering
\includegraphics[width=0.5\textwidth]{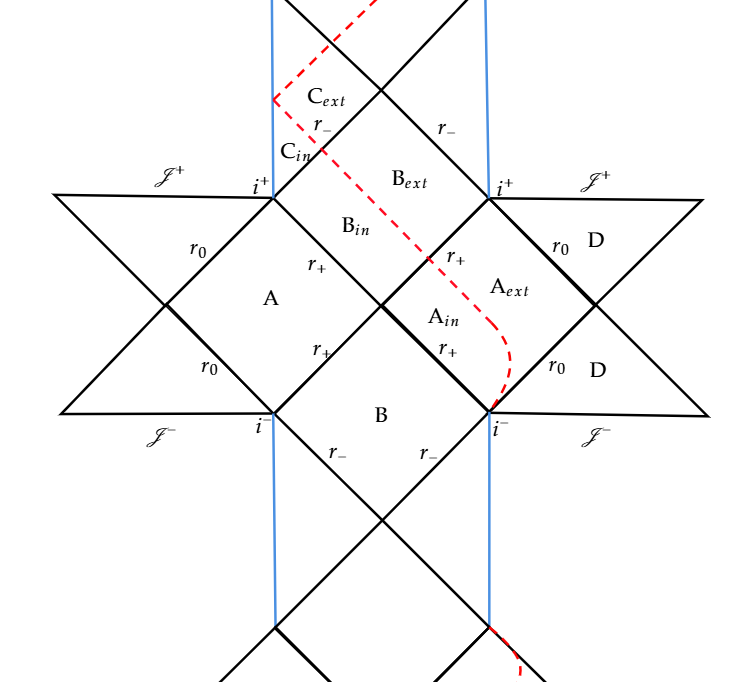}
     
\caption{Penrose diagram for the maximum extension of the spacetime with three horizons in the exterior area of the dust bull. However, only the areas $\mathrm{A}_{ext}$, $\mathrm{B}_{ext}$ and $\mathrm{C}_{ext}$ and $\mathrm{D}$ are real and the metric (\ref{metric}) can make sense.}
        \label{fig:000}
\end{figure}

But in our paper, we focus on the exterior spacetime of the qOS-dS black hole whose metric is given by~\cite{PhysRevD.109.064012,Lin:2024flv}:
\begin{eqnarray}\label{metric}
	\mathrm{d}s^2_{ex}=-f(r)\mathrm{d}t^2+\frac{\mathrm{d}r^2}{f(r)}+r^2(\mathrm{d}\theta^2+\sin^2\theta\mathrm{d}\phi^2)\,,
\end{eqnarray}
where the metric function $f(r)$ reads
\begin{eqnarray}\label{metric_function}
	f(r)=1-\frac{2M}{r}-\frac{\Lambda}{3}r^2+\frac{\alpha M^2}{r^4}\Big(1+\frac{\Lambda r^3}{6M}\Big)^2\,,
\end{eqnarray}
with the positive parameter $\alpha=16\sqrt{3}\pi\gamma^3l^2_p$.

The most realistic case is when there is three horizons, corresponding to three positive roots of the metric. If we set $r_{-}$, $r_{+}$ and $r_{0}$ in response to the inner, outer and cosmological horizon respectively, then one special case is that when $\Lambda=0$ and the spacetime is asymptotically flat, then there are only two horizons expected, under which following the method in~\cite{Cao:2024oud} and setting $0<q^2=r_{-}/r_{+}<1$ one can easily get the relation
\begin{eqnarray}
     \frac{\alpha}{M^2}=\frac{16q^6(1+q^2+q^4)^3}{(1+q^2+q^4+q^6)^4}\,,
\end{eqnarray}
and 
\begin{eqnarray}
    r_{+}=2M\frac{1+q^2+q^4}{1+q^2+q^4+q^6}\,,\quad
    r_{-}=2Mq^2\frac{1+q^2+q^4}{1+q^2+q^4+q^6}\,,
\end{eqnarray}
and the more general case is that there is also a cosmological horizon, for which through similar strategy, that is, setting $0<q^2=r_{-}/r_{+}<1$
 and $0<p^2=r_{+}/{r_{0}}<1$,
then the metric can be rewritten as 
\begin{eqnarray}\label{oo}
    f(r)=\Big(\frac{\alpha \Lambda^2}{36}-\frac{\Lambda}{3}\Big)r^2\Big(1-\frac{r_{0}}{r}\Big)\Big(1-\frac{p^2r_{0}}{r}\Big)\Big(1-\frac{p^2q^2r_{0}}{r}\Big)\Big(1+\frac{dr_{0}}{r}+\frac{br_{0}^2}{r^2}+\frac{cr_{0}^3}{r^3}\Big)\,,
\end{eqnarray}
where
\begin{eqnarray}
    d=1+p^2+q^2p^2\,,\quad
    b=\frac{p^2q^2(1+p^2+q^2p^2)(1+q^2+q^2p^2)}{1+q^2+q^4+p^2q^2+p^2q^4+p^4q^4}\,,\quad
     c=\frac{p^4q^4(1+p^2+q^2p^2)}{1+q^2+q^4+p^2q^2+p^2q^4+p^4q^4}\,,
\end{eqnarray}
in this way one is easy to find the relation of parameters including
\begin{eqnarray}\label{0}
   \alpha M^2r_{0}^{-4}=\tilde{a}\tilde{c}\, ,\quad
   \Big(2-\frac{\alpha \Lambda}{3}\Big)\frac{M}{r_{0}}=\tilde{b}\tilde{c}\, ,\quad
   \Big(\frac{\Lambda}{3}-\frac{\alpha \Lambda^2}{36}\Big)r_{0}^2=\tilde{c}\, ,
\end{eqnarray}
where
\begin{eqnarray}
    \tilde{a}=\frac{p^8q^6(1+p^2+q^2p^2)}{1+q^2+q^4+p^2q^2+p^2q^4+p^4q^4}\,,
\end{eqnarray}
\begin{eqnarray}
    \tilde{b}=p^2(1+p^2)(1+q^2)\frac{1+q^4+p^4q^4+p^6q^6+p^2(q^2+q^6)}{1+q^2+q^4+p^2q^2+p^2q^4+p^4q^4}\,,
\end{eqnarray}
\begin{eqnarray}
    \tilde{c}=\frac{1+q^2+q^4+p^2q^2+p^2q^4+p^4q^4}{(1+p^2+p^4)(1+q^2+q^4)(1+p^2q^2+p^4q^4)}\, .
\end{eqnarray}
By combing the relations in Eqs.(\ref{0}), one is able to get more explicit relations between given parameters: 
\begin{eqnarray}
    \Lambda \alpha=6\Big(1- \sqrt{\frac{1}{1+4\tilde{a}/\tilde{b}^2}}\Big)\,,\quad
    \frac{\alpha}{M^2}=\frac{16\tilde{a}}{\tilde{c}^3(\tilde{b}^2+4\tilde{a})^2}\, ,\quad
    \frac{r_{0}}{M}=\frac{2}{\tilde{c}}\sqrt{\frac{1}{\tilde{b}^2+4\tilde{a}}}\, .
\end{eqnarray}

\begin{figure}[htbp]
	\centering
\includegraphics[width=0.45\textwidth]{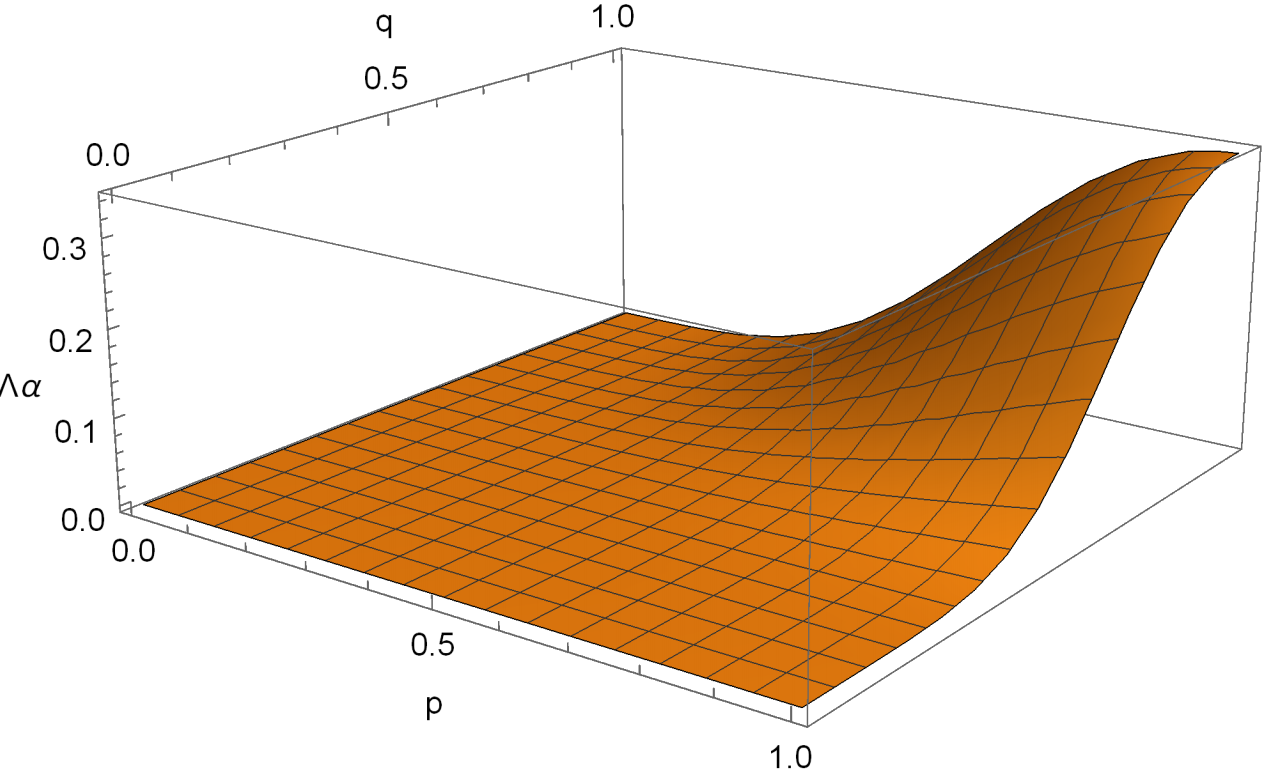}
     \includegraphics[width=0.45\textwidth]{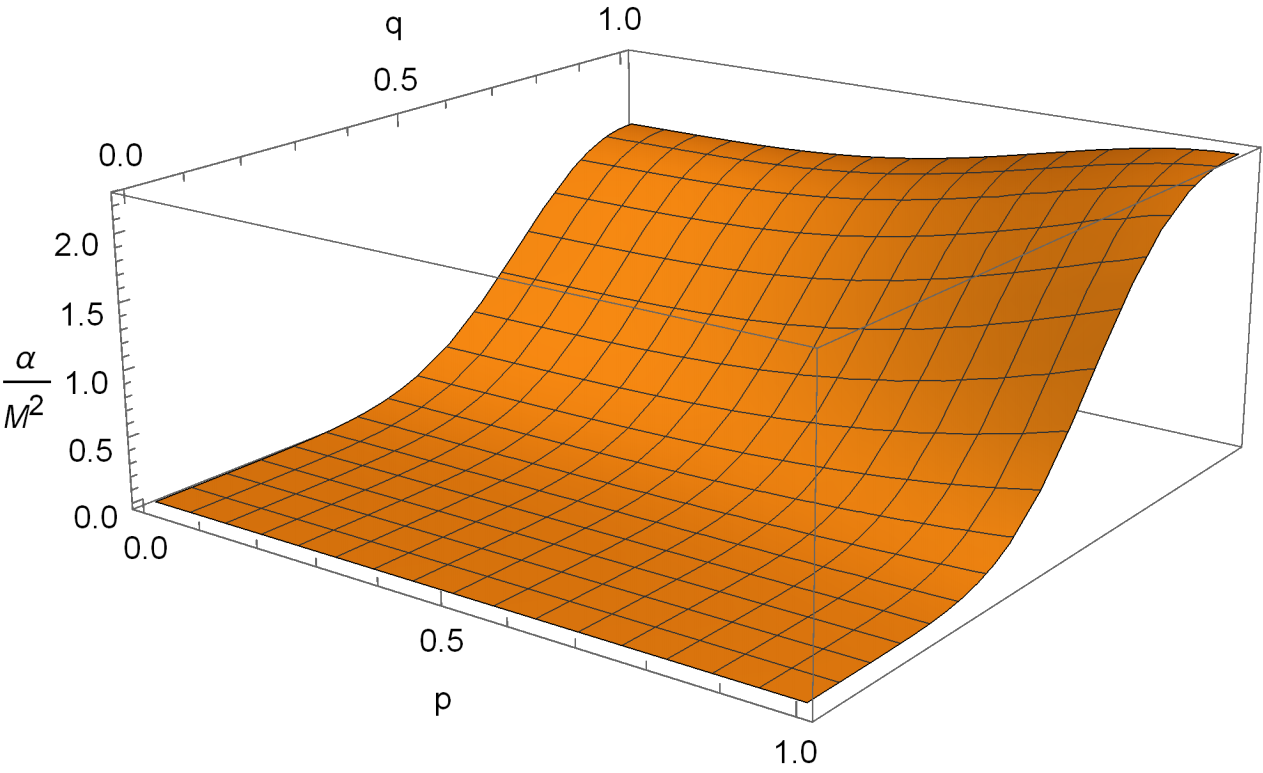}   \\ 
     \includegraphics[width=0.45\textwidth]{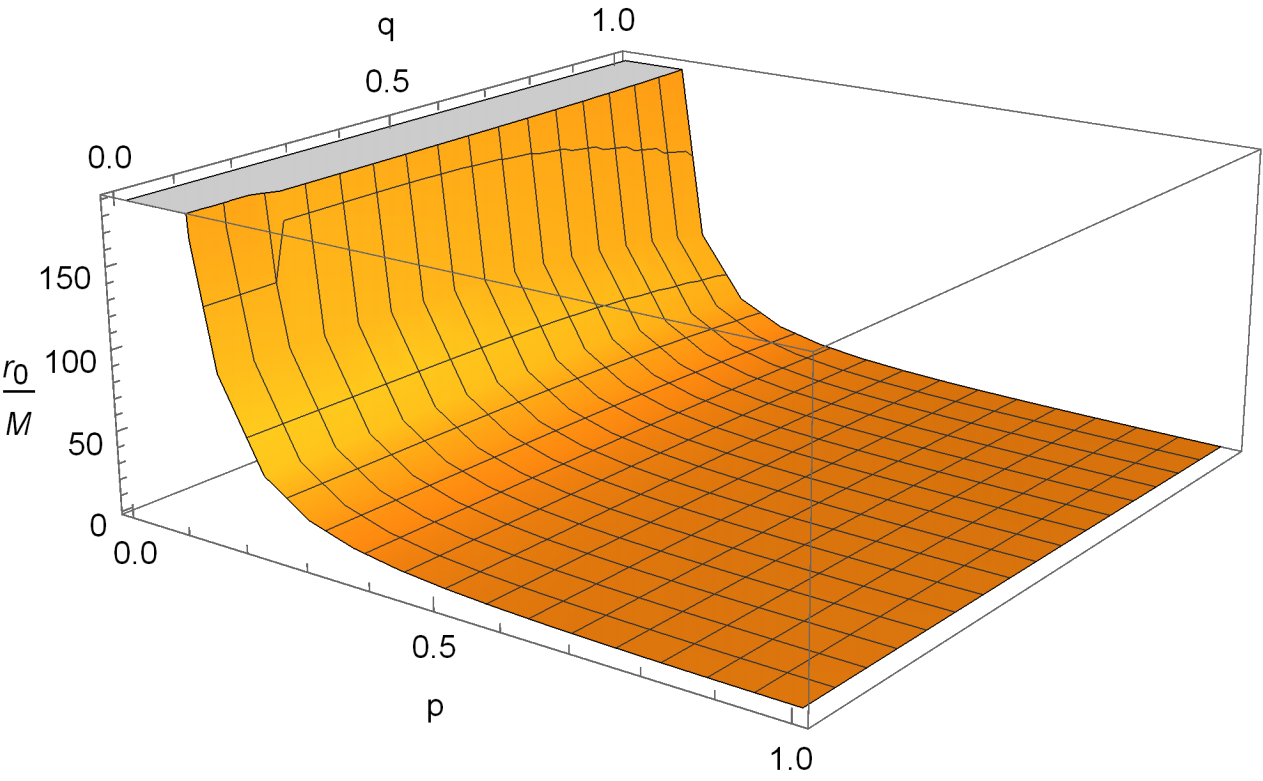}
     \includegraphics[width=0.45\textwidth]{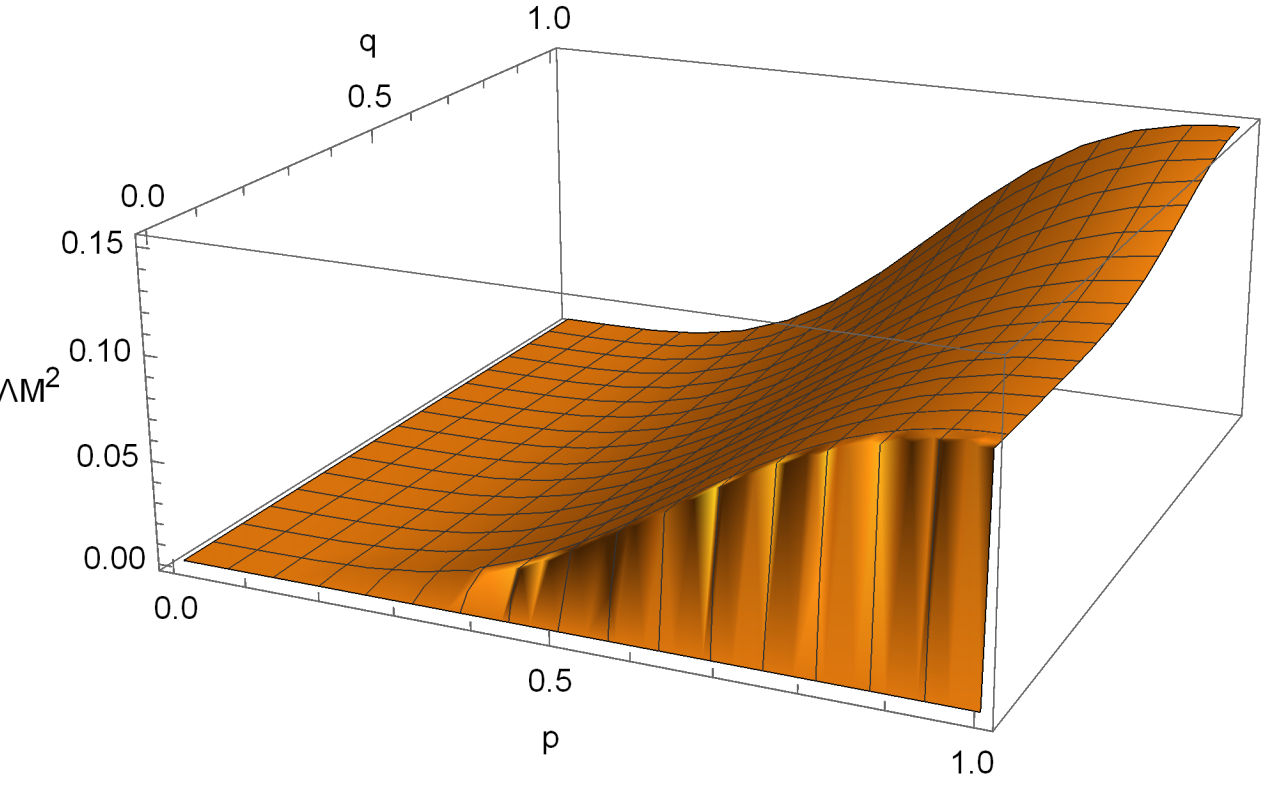}    
\caption{The relation between some given parameters and the relative ratio of the three horizon radius.  }
        \label{fig:001}
\end{figure}

From Fig.\ref{fig:001}, it is shown that the largest value of $\Lambda \alpha$ is $6-4\sqrt{2}$, if and only if $p^2=q^2=1$, i.e., the three horizons merge. At the same time, we have $\alpha/M^2=9/4$ and $r_{0}/M=3\sqrt{2}/2$ as $p^2=q^2=1$. Through comparing the top right panel and the bottom right panel, one is able to find as long as the black hole mass $M$ is fixed, $p$ mostly relies on parameter $\Lambda$, while $q$ mostly relies on the parameter $\alpha$. That is a natural conclusion considering the physical meaning of these two factors. The main advantage of this parameterization is the convenience in the following numerical calculation, in which extreme case is easier to study.

For the details of this spacetime, we can effectively adopt the semi-classical method and suppose the metric of the spacetime still obey the Einstein equation with nothing but a nonzero quantum energy-momentum tensor~\cite{PhysRevLett.130.101501}. Now one can easily calculate the non-zero components of the energy momentum tensor $T_{\mu\nu}$ as
\begin{eqnarray}
    T_{00}=\frac{(36\alpha M^2+12\Lambda r^6-\alpha\Lambda^2r^6)[\alpha(6M+\Lambda r^3)^2-12r^3(6M-3r+\Lambda r^3)]}{432r^{10}}\\+\Lambda \Bigg{(}1-\frac{2M}{r}-\frac{\Lambda}{3}r^2+\frac{\alpha M^2}{r^4}\Big(1+\frac{\Lambda r^3}{6M}\Big)^2\Bigg{)}\,,\qquad \qquad \qquad
\end{eqnarray}
\begin{eqnarray}
    T_{11}=\frac{3\Lambda(\alpha\Lambda-12)r^6-108\alpha M^2}{\alpha(6Mr+\alpha r^4)^2-12r^5(6M-3r+\Lambda r^3)}+\frac{\Lambda}{1-\frac{2M}{r}-\frac{\Lambda}{3}r^2+\frac{\alpha M^2}{r^4}\Big(1+\frac{\Lambda r^3}{6M}\Big)^2}\,,
\end{eqnarray}
\begin{eqnarray}
    T_{22}=\frac{6\alpha M^2}{r^4}+\frac{\Lambda^2}{12}\alpha\,,\qquad T_{33}=\frac{[72\alpha M^2+\Lambda^2\alpha r^6]\mathrm{sin}^2\theta}{12r^4}+\Lambda r^2\mathrm{sin}^2\theta\,,
\end{eqnarray}
and some of important scalars include 
\begin{eqnarray}
  R=4\Lambda-\frac{\alpha \Lambda^2}{3}-\frac{6M^2\alpha}{r^6}\,,  
\end{eqnarray}
\begin{eqnarray}
  R_{ab}R^{ab}=\frac{90M^4\alpha^2}{r^{12}}+\frac{M^2\alpha\Lambda(\alpha\Lambda-12)}{r^6}+\frac{\Lambda^2(\alpha\Lambda-12)^2}{36}\,,
\end{eqnarray}
\begin{eqnarray}
R_{abcd}R^{abcd}=\frac{468M^4\alpha^2}{r^{12}}+\frac{40M^3\alpha(\alpha\Lambda-6)}{r^9}+\frac{2M^2(\alpha^2\Lambda^2-12\alpha\Lambda+24)}{r^6}+\frac{\Lambda^2(\alpha\Lambda-12)^2}{54}\,.
\end{eqnarray}
 Here we need to emphasize that the work above is only an effective treatment of quantum effect, but with no more physical meaning, especially with the test Proca field. For Proca field, if we ignore the quantum effect, which is rational at least in area far away from the possible smallest radius of the dust ball, we can still regard the vector field to evolve freely with no coupling to the spacetime structure or gravity perturbation.

\section{Axial Proca field Perturbation of qOS-dS model}\label{sec:2}

In this section, we will study the perturbation of the Proca field in the qOS-dS spacetime. The premise of our study is the semi-classical approximation, that is, to regard the Proca field as a classical field on a spacetime metric influenced by quantum effect. Following the method proposed in~\cite{Chad1984}, for the spacetime described by   
\begin{eqnarray}
    \mathrm{d}s^2=-f(r)\mathrm{d}t^2+\frac{\mathrm{d}r^2}{g(r)}+r^2(\mathrm{d}\theta^2+\mathrm{sin}^2\theta\mathrm{d}\phi^2)\,,
\end{eqnarray}
 we study the linear Einstein-Proca equation in an orthonormal frames (in the following calculation we use $f$ to stand for $f(r)$ and $g$ to stand for $g(r)$):
\begin{eqnarray}\label{1099}
    e_{(0)\mu}=(-\sqrt{f},0,0,0)\,,\qquad
    e_{(1)\mu}=(0,\frac{1}{\sqrt{g}},0,0)\,,\qquad
    e_{(2)\mu}=(0,0,r,0)\,,\qquad
    e_{(3)\mu}=(0,0,0,r\mathrm{sin}\theta)\,.
\end{eqnarray}
or, equivalently, 
\begin{eqnarray}
    e_{(0)}^{\mu}=(\frac{1}{\sqrt{f}},0,0,0)\,,\quad
    e_{(1)}^{\mu}=(0,\sqrt{g},0,0)\,,\quad
    e_{(2)}^{\mu}=(0,0,\frac{1}{r},0)\,,\quad
    e_{(3)}^{\mu}=(0,0,0,\frac{1}{r\mathrm{sin}\theta})\,. 
\end{eqnarray}
in which later calculation shows obvious simplification in the equation's expression, as there is $e_{(a)}^{\mu}e_{(b)\mu}=\eta_{(a)(b)}$. 
Now we turn to the study of the axial perturbation of Proca field. ``Axial" means the rotation of the field around a certain axis, which contributes to an only nonzero component of vector potential in $A_{3}$, and this component can contribute to nonzero $F_{03},F_{13}$ and $F_{23}$. If the original $A_{0}$ is nonzero, then this perturbation can also contribute to the non diagonal metric components including $g_{03},g_{13},g_{23}$ but no other components. It can be proved that this kind of perturbation is independent of another perturbation called polar perturbation~\cite{Chad1984}. In the frame introduced earlier, the Proca equation has the form of 
\begin{eqnarray}
    \eta^{(m)(n)}F_{(a)(m)|(n)}+m^2A_{(a)}=0\,,\qquad
    F_{[(a)(m)|(n)]}=0\,,
\end{eqnarray}
However, as the specific expression of this equation requires the calculation of the rotation coefficients, we firstly adopt the common Proca equation in the coordinates:
\begin{eqnarray}  \nabla_{\mu}F^{\mu\nu}+m^2A_{\nu}=0\,,\qquad
   \nabla_{[\mu}F_{\nu\rho]}=0\,,  
\end{eqnarray}
or, explicitly, for axial perturbation: 
\begin{eqnarray}
    \sqrt{\frac{g}{f}}\frac{1}{r^2}\frac{\partial}{\partial r}( \sqrt{\frac{f}{g}}r^2\mathrm{sin}\theta F^{13})+\frac{\partial}{\partial\theta}(\mathrm{sin}\theta F^{23})+\frac{\partial}{\partial t}(\mathrm{sin}\theta F^{03})+m^2A^{3}=0\,,
\end{eqnarray}
together with
\begin{eqnarray}    \partial_{r}F_{03}+\partial_{t}F_{31}=0\,,\qquad
\partial_{\theta}F_{03}+\partial_{t}F_{32}=0\,,  
\end{eqnarray}
Considering the relation 
\begin{eqnarray}
    F^{\mu\nu}=F^{(a)(b)}e_{(a)}^{\mu}e_{(b)}^{\nu}=\eta^{(c)(a)}\eta^{(d)(b)}F_{(c)(d)}e_{(a)}^{\mu}e_{(b)}^{\nu}\,,
\end{eqnarray}
thus all we need are
\begin{eqnarray}\label{1}
    m^2\sqrt{\frac{f}{g}}rA_{(3)}+\frac{r}{\sqrt{g}}F_{(0)(3),t}+(r\sqrt{f}F_{(3)(1)})_{,r}+\sqrt{\frac{f}{g}}F_{(3)(2),\theta}=0\,,
\end{eqnarray}
\begin{eqnarray}\label{2} (r\sqrt{f}F_{(0)(3)}\mathrm{sin}\theta)_{,r}+\frac{r}{\sqrt{g}}F_{(3)(1),t}\mathrm{sin}\theta=0\,,\qquad
    r\sqrt{f}(F_{(0)(3)}\mathrm{sin}\theta)_{,\theta}+r^2F_{(3)(2),t}\mathrm{sin}\theta=0\,,
\end{eqnarray}
insert (\ref{2}) into (\ref{1}) and note that $A_{(3),t}=\sqrt{f}F_{(0)(3)}$ we get 

\begin{eqnarray}
    r[\frac{\Omega^2}{\sqrt{g}}-m^2\frac{f}{\sqrt{g}}]F_{(0)(3)}+\frac{f}{\sqrt{g}r}\frac{\partial}{\partial\theta}[\mathrm{sin}\theta\frac{\partial}{\partial\theta}(F_{(0)(3)}\mathrm{sin}\theta)]+\frac{\partial}{\partial r}[\sqrt{fg}\frac{\partial}{\partial r}(r\sqrt{f}F_{(0)(3)})]=0\,.
\end{eqnarray} 
here we have supposed a feature frequency $\omega$ to replace the derivative to $t$. Then by variable separating we get
\begin{eqnarray}\label{10}
     \frac{\mathrm{d}}{\mathrm{d}r}[\sqrt{fg}\frac{\mathrm{d}}{\mathrm{d}r}(r\sqrt{f}B)]-l(l+1)\frac{fB}{\sqrt{g}r}+r(\frac{\omega^2}{\sqrt{g}}-m^2\frac{f}{\sqrt{g}})B=0\,.
\end{eqnarray}
where $B$ is the radial part of the whole perturbation, $l$ is the angular momentum number.
 
\section{Numerical approach to calculate QNMs of axial Proca field perturbations in qOS-dS Black Hole}\label{sec:3}

\begin{figure}[htbp]
	\centering
\includegraphics[width=0.4\textwidth]{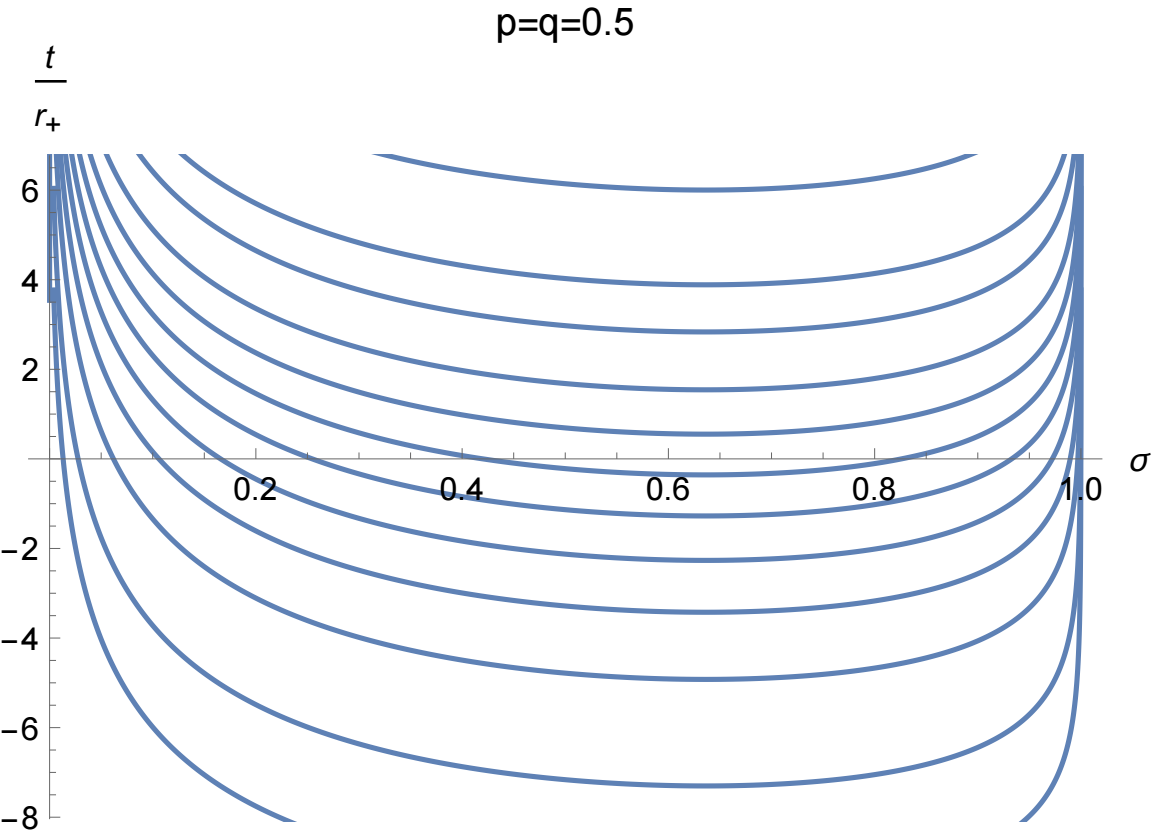}

        \caption{The $\tau=$constant hypersurfaces in $t-\sigma$ diagram under the parameters $p=q=0.5$. It's shown that while $\tau$ remains constant $t$ approaches infinity on both sides.}
        \label{fig:dia}
\end{figure}

Firstly, we should make it clear that for a non vanishing $\Lambda$ the region we study is between the outer horizon ($r_{+}$) and cosmological horizon ($r_{0}$) of the qOS model. Under this basic assumptions, there is always $V(r)\rightarrow 0$ no matter for $r\rightarrow r_{+}$ or $r\rightarrow r_{0}$, so both sides are null boundary condition, and more specifically, there is $\Phi\sim e^{i\omega (t- r_{*})}$ near $r=r_{0}$ and $\Phi\sim e^{i\omega (t+ r_{*})}$ near $r=r_{+}$, which accords to conventional definition for QNMs, even though here the field is massive. This offers the central prerequisite for using hyperboloidal approach, which are fundamentally revealed in~\cite{Zengino_lu_2008}. The hyperboloidal approach is able to absorb the boundary asymptotic condition into the non-divergence condition of the differential operator near the boundary, which is naturally satisfied in discretizing and matrix approximation. However, for $\Lambda=0$ and an asymptotically flat spacetime, the boundary condition is time-like for massive fields, which cause the traditional hyperbolodial methods in calculating QNMs lose effectiveness. Here we avoid to discuss this condition. For the special case $f=g$, we set
\begin{eqnarray}
    \Phi=\frac{-r\sqrt{f}B}{2\sqrt{n}}\,,\qquad
    \mathrm{d}r_{*}=\frac{\mathrm{d}r}{f}\,,
\end{eqnarray}
and insert into (\ref{10}), we get the wave equation
\begin{eqnarray}\label{09090}
   \Big{[}\frac{\mathrm{\partial}^2}{\partial r_{*}^{2}}+\omega^2-\frac{f}{r}\Big{(}\frac{l(l+1)}{r}+m^2r\Big{)}\Big{]} \Phi=0\,, \label{we}
\end{eqnarray}
 When $m=0$, the potential is exactly the Regge-Wheeler potential for $s=1$. As the space region we study is $r_{+}<r<r_{0}$, using the minimal gauge of hyperboloidal we transform the coordinate to 

\begin{eqnarray}\label{ws}
    r_{*}=2r_{+}h_{1}(\sigma)\,,\qquad
    t=2r_{+}(\tau-h_{2}(\sigma))\,,
\end{eqnarray}
using the minimal gauge, the explicit form of the transformation can be written as
 \begin{eqnarray}\label{00}
    \sigma=\frac{{r}_{+}r_{0}}{(r_{0}-r_{+})r}-\frac{r_{+}}{r_{0}-r_{+}}\,, \qquad
\mathrm{d}\tau=\frac{\mathrm{d}t-\mathrm{d}r_{*}}{2{r}_{+}}+\frac{\beta r_{+}\mathrm{d}r}{r(r-{r}_{+})}\,,
\end{eqnarray}
where 
\begin{eqnarray}\label{beta}
    \beta=-\frac{1-p^2}{f'(1)}\,,
\end{eqnarray}
in which $p^2=r_{+}/r_{0}$ and $f$ is written as a function of $\sigma$. Fig.\ref{fig:dia} shows the effect of this transformation: the original infinity in $t-r$ domain is drawn closer and divided into several $\tau=$constant hypersurfaces.

Following the method in~\cite{PhysRevX.11.031003,Cao:2024oud}, we set
$\Psi=\partial_{\tau} \Phi$, then in $(\sigma,\tau)$ coordinate there is
\begin{eqnarray}\label{12345}
i\omega\begin{bmatrix}
        \Phi\\
        \Psi
    \end{bmatrix}=
\partial_{\tau}
  \begin{bmatrix}
        \Phi\\
        \Psi
    \end{bmatrix}
    =\begin{bmatrix}
        O& 1\\
        L_{1} & L_{2}
    \end{bmatrix}
     \begin{bmatrix}
        \Phi\\
        \Psi
    \end{bmatrix}\equiv\textbf{L}\begin{bmatrix}
        \Phi\\
        \Psi
    \end{bmatrix}\,,
\end{eqnarray}
where 
\begin{eqnarray}\label{19}
    L_{1}=\frac{1}{w_{1}(\sigma)}(\partial_{\sigma}(\nu(\sigma)\partial_{\sigma})-Q(\sigma))\,,
\end{eqnarray}
\begin{eqnarray}\label{18}
    L_{2}=\frac{1}{w(\sigma)}(2\gamma_{1}(\sigma)\partial_{\sigma}+\partial_{\sigma}\gamma(\sigma))\,,
\end{eqnarray}
and
\begin{eqnarray}\label{198}
\nu(\sigma)=\frac{2[(1-p^2)\sigma+p^2]^2f}{1-p^2}\,, \qquad
Q(\sigma)=\frac{2{r}_{+}^2(1-p^2)V}{[(1-p^2)\sigma+p^2]^2f} \,,
\end{eqnarray}
\begin{eqnarray}\label{194}
  w(\sigma)=\frac{2\beta}{1-\sigma}-\frac{2\beta^2[(1-p^2)\sigma+p^2]^2f}{(1-\sigma)^2(1-p^2)}\,,\qquad
\gamma(\sigma)=1-\frac{2\beta f[(1-p^2)\sigma+p^2]^2}{(1-p^2)(1-\sigma)} \,. 
\end{eqnarray}
and here $\omega$ actually stands for dimensionless $2r_{+}\omega$.

\begin{figure}[htbp]
	\centering
\includegraphics[width=0.4\textwidth]{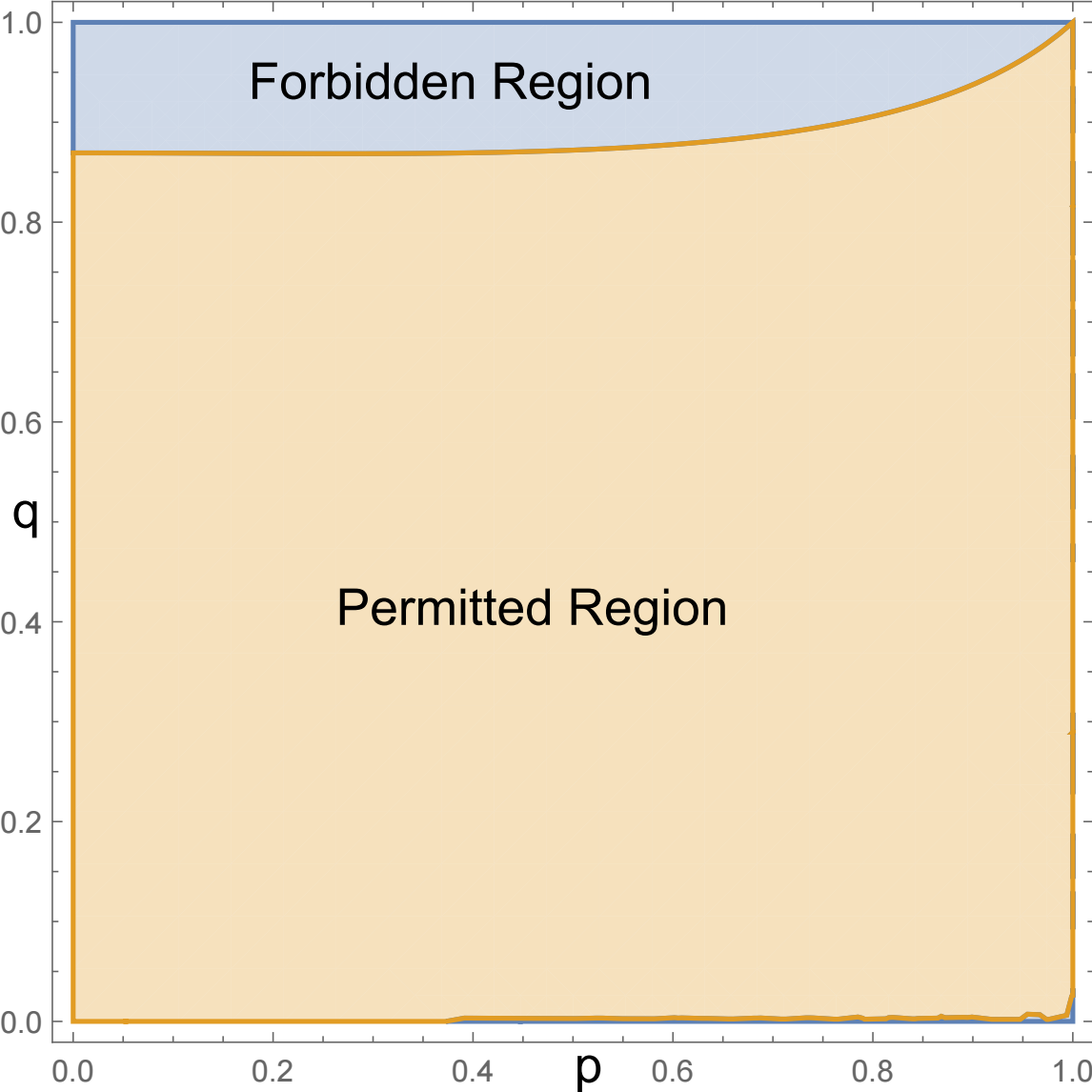}

        \caption{The $p-q$ diagram for the model, where the orange region corresponds to the permitted region for our method while the blue region corresponds to the forbidden region.}
        \label{fig:fp}
\end{figure}

However, the effectiveness of the method in Eqs.(\ref{198}) and Eqs.(\ref{194}) relies on the non-divergence of the elements in the calculating matrix in Eq.(\ref{12345}). Although the setting of the coefficient in Eq.(\ref{beta}) have already precisely guaranteed that $w(\sigma), \gamma(\sigma)$ and their reciprocals are regular on the two edges and the vanishing of the effective potential on the edges guaranteed the non-divergence of 
$\textbf{Q}(\sigma)$, there is still a residual danger that under certain parameters whether $w(\sigma)$ could have a root between $\sigma=0$ and $\sigma=1$. Unfortunately for qOS-dS model there really exists such a forbidden region of parameters $p$ and $q$ where our hyperboloidal coordinates construction loses effectiveness, which is shown in Fig.\ref{fig:fp}. 

So what happens when the parameters are in the forbidden region? Actually one is easy to confirm that according to the general transformation rule in Eqs.(\ref{ws}) the hypersurface $\tau=$constant is marked by its normal vector $(\mathrm{d}\tau)_{a}$, and there is
\begin{eqnarray}
    g^{ab}(\mathrm{d}\tau)_{a}(\mathrm{d}\tau)_{b}=-\frac{1}{(2r_{+})^2f}\frac{h_{1}'^2-h_{2}'^2}{h_{1}'^2-h_{2}'^2+h_{1}'^2h_{2}'^2}\,,
\end{eqnarray}
in which $h_{1}'$ and $h_{2}'$ stands for the derivatives to coordinate $\sigma$. Meanwhile there is 
\begin{eqnarray}
    w(\sigma)=\frac{h_{1}'^2-h_{2}'^2}{|h_{1}'|}\,,
\end{eqnarray}

Usually there is $g^{ab}(\mathrm{d}\tau)_{a}(\mathrm{d}\tau)_{b}<0$, which means the $\tau=$constant surface is a space-like one and accords to the demand, however, this is not always true. It might be guaranteed on one side (in our method $\sigma=0$), when being violated on the other, which, according to the continuity, must lead to one root of $g^{\tau\tau}$ as well as the function $w(\sigma)$ between $\sigma=0$ and $\sigma=1$. Under this condition our coordinate hypersurface turns out to be space-like only on partial $\sigma$ domain but null and even time-like on the others. And in the previous studies although the asymptotic behavior of $\tau$ coordinate line on the two horizons is emphasized for many times, this fact is often neglected. Although this is a meaningful warning for coordinate constructing, here for convenience we just avoid talking about the forbidden region or using more complex hyperboloidal gauge.

To calculate the QNMs, we firstly transfer
 the variable range from $(0,1)$ to $(-1,1)$:
\begin{eqnarray}
    x=2\sigma-1 \,,
\end{eqnarray}
we sample $N+1$ points from a Chebyshev-Gauss-Labatto grid in which the points are the extrema of Chebyshev polynomial on the interval $(-1,1)$, i.e.,
\begin{eqnarray}\label{Chebyshev_Gauss_Labatto_grid}
    x_{j}=\cos\Big(\frac{j\pi}{N}\Big)\, ,\quad j=0,1,\cdots,N\,,
\end{eqnarray}
and follow~\cite{PhysRevX.11.031003} we can discretize the original derivative matrix equation (\ref{12345}) to 
\begin{eqnarray}\label{ooo}
    i\omega\begin{bmatrix}
        \mathbf{\Phi}\\
        \mathbf{\Psi}
    \end{bmatrix}=
     \textbf{L}\begin{bmatrix}
       \mathbf{\Phi}\\
        \mathbf{\Psi}
    \end{bmatrix}=
    \begin{bmatrix}
        \textbf{O} &\textbf{I}_{N}\\
        \textbf{L}_{1}^{(2)}\textbf{D}_{N}^2+ \textbf{L}_{1}^{(1)}\textbf{D}_{N}+\textbf{V}&\textbf{L}_{2}^{(1)}\textbf{D}_{N}+\textbf{L}_{2}^{(0)}\\
    \end{bmatrix}
 \begin{bmatrix}
        \mathbf{\Phi}\\
        \mathbf{\Psi}
    \end{bmatrix}\,,
\end{eqnarray}
where 
\begin{eqnarray}
    \mathbf{\Phi}=(\Phi(x_0),\Phi(x_1),\cdots,\Phi(x_{N-1}),\Phi(x_N))^T\,,
\end{eqnarray}
\begin{eqnarray}
    \mathbf{\Psi}=(\Psi(x_0),\Psi(x_1),\cdots,\Psi(x_{N-1}),\Psi(x_N))^T\,,
\end{eqnarray}
and
\begin{eqnarray}\label{Chebyshev_differential_matrix}
    (\textbf{D}_N)_{00}=\frac{2N^2+1}{6}\, ,\quad (\textbf{D}_N)_{NN}=-\frac{2N^2+1}{6}\, ,\nonumber\\
    (\textbf{D}_N)_{jj}=\frac{-x_j}{2(1-x_j^2)}\, ,\quad j=1\,\cdots,N-1\, ,\nonumber\\
    (\textbf{D}_N)_{ij}=\frac{c_i}{c_j}\frac{(-1)^{i+j}}{(x_i-x_j)}\, ,\quad i\neq j\, ,\quad i,j=0,\cdots,N\, ,\\
\end{eqnarray}
\begin{eqnarray}\label{c_i}
	c_i=\left\{
	\begin{aligned}
		&2&\, ,\quad i=0\, \text{or}\, N\, ,\\
		&1&\, ,\quad \text{otherwise}\, .
	\end{aligned}
	\right. 
\end{eqnarray}
and
\begin{eqnarray}
    (\textbf{L}_{1}^{(2)})_{ij}=\frac{4\alpha(\frac{x_{i}+1}{2})\delta_{ij}}{w(\frac{x_{i}+1}{2})},\qquad
    (\textbf{L}_{1}^{(1)})_{ij}=\frac{4\partial_{x}\alpha(\frac{x_{i}+1}{2})\delta_{ij}}{w(\frac{x_{i}+1}{2})}\,,\\
     (\textbf{L}_{2}^{(1)})_{ij}=\frac{4\gamma(\frac{x_{i}+1}{2})\delta_{ij}}{w(\frac{x_{i}+1}{2})}\,,\qquad
     (\textbf{L}_{2}^{(0)})_{ij}=\frac{2\partial_{x}\gamma(\frac{x_{i}+1}{2})\delta_{ij}}{w(\frac{x_{i}+1}{2})}\,,\\
      (\textbf{V})_{ij}=\frac{Q(\frac{x_{i}+1}{2})\delta_{ij}}{w(\frac{x_{i}+1}{2})}\,.\qquad \qquad \qquad
\end{eqnarray}

\section{Instability of the QNMs}\label{sec:41}
\subsection{Parameter Perturbation}\label{sec:10}
About the stability under a certain perturbation, we introduce the migration susceptibility defined as
\begin{eqnarray}\label{delta_omega}
\chi_{n}\equiv\displaystyle\lim_{\epsilon\rightarrow 0} \frac{\delta \omega_{n}}{\epsilon}\, ,
\end{eqnarray}
to quantify the relative instability of a certain QNM, where $n$ stands for the overtone number and $\delta \omega_{n}$ refers to the change. The overtone is defined according to the imaginary part absolute value, with the fundamental mode considered as the mode with the smallest imaginary part. 

For the most general discussion, we replace $r_{*}$ by $x$, and suppose the original potential to be $V(x)$ with the migration to be $\epsilon V_{b}(x)$. The definition of Jost solutions is those solutions of Eq.(\ref{we}) satisfying the conditions~\cite{yang2024spectralinstabilityblackholes}
\begin{eqnarray}\label{f1}
    \phi_{\omega}(x)\rightarrow \mathrm{e}^{i\omega x}\Big{[}1+O(\frac{1}{x})\Big{]}\,, \qquad x\rightarrow -\infty \label{2001}
\end{eqnarray}
\begin{eqnarray}
        \psi_{\omega}(x)\rightarrow \mathrm{e}^{-i\omega x}\Big{[}1+O(\frac{1}{x})\Big{]}\,, \qquad x\rightarrow +\infty \label{2002}
\end{eqnarray}
and as there are only two linear independent solutions to wave equation, there is the relation
\begin{eqnarray}\label{e1}
    \phi_{\omega}(x)=a(\omega)\psi_{-\omega}(x)+b(\omega)\psi_{\omega}(x)\,, 
\end{eqnarray}
and through this definition it's easy to see all the QNMs are the roots of the $a(\omega)$, so considering the perturbation from $a^{(0)}(\omega)$ to $a^{(\epsilon)}(\omega)$ caused by $\epsilon V_{b}(x)$, there is the relation 
\begin{eqnarray}
    0=a^{(0)}(\omega_{n})=a^{(\epsilon)}(\omega^{(\epsilon)}_{n})\approx a^{(\epsilon)}(\omega_{n})+a^{(\epsilon)'}(\omega_{n})\delta\omega_{n}\,,
\end{eqnarray}
so there is 
\begin{eqnarray}
    \chi_{n}=-\displaystyle\lim_{\epsilon\rightarrow 0}\Bigg[\frac{1}{\epsilon}\frac{a^{(\epsilon)}(\omega_{n})}{a^{(0)'}(\omega_{n})}\Bigg]\,,
\end{eqnarray}
and from the definition in (\ref{2001}) (\ref{2002}), one is able to solve the explicit form of the coefficient $a(\omega)$:
\begin{eqnarray}
    a(\omega)=1-\int^{\infty}_{-\infty}\frac{\mathrm{e}^{i\omega y}}{2i\omega}V(y)\phi_{\omega}(y)\mathrm{d}y\,,
\end{eqnarray}
and the numerator of $\chi_{n}$ can be approximated as
\begin{eqnarray}
    a^{(\epsilon)}(\omega_{n})\approx -\epsilon \int^{\infty}_{-\infty}\frac{\mathrm{e}^{i\omega_{n} y}}{2i\omega_{n}}V_{b}(y)\phi_{\omega_{n}}(y)\mathrm{d}y\,,
\end{eqnarray}
and is usually a regular number compared to the order $\epsilon$ as long as $V_{b}$ is also a short-ranged and positive potential so it does not contribute much to either stability or the instability. About the denominator, however, things are quite different, as there is 
\begin{eqnarray}
    a^{(0)'}(\omega_{n})=-\int^{\infty}_{-\infty}\frac{\mathrm{e}^{i\omega_{n} y}}{2i\omega_{n}}V(y)\Big[\frac{i\omega_{n}y-1}{\omega_{n}}\phi_{\omega_{n}}(y)+\frac{\delta\phi_{\omega_{n}}(y)}{\delta\omega_{n}}]\mathrm{d}y\,,
\end{eqnarray}
and if this denominator turns to be 0, then there is a so-called Type I instability \cite{yang2024spectralinstabilityblackholes}, in which $\chi_{n}$ under a perturbation turns to be singular, and it's clear that such singularity has nothing to do with the specific form of perturbation.

Now that we have constructed the hyperboloidal framework in Sec.\ref{sec:3}, we can use the calculation results of the first overtones to study the instability and their response to parameter migration. We take $N=48$ as the dimension of the discrete matrix $\textbf{L}$ and machine precision to be 120 and we confirm that at least for the first four overtones the results are convergent as $N$ increases. 

\begin{figure}[htbp]
\includegraphics[width=0.43\textwidth]{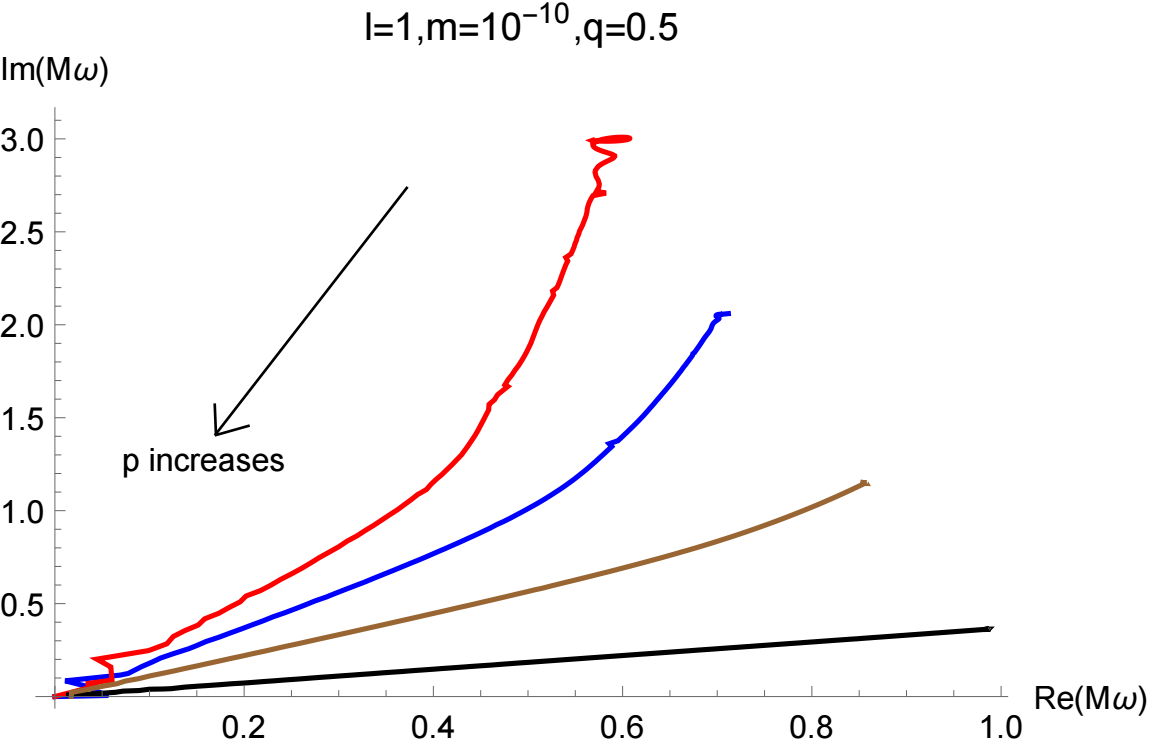}
\includegraphics[width=0.43\textwidth]{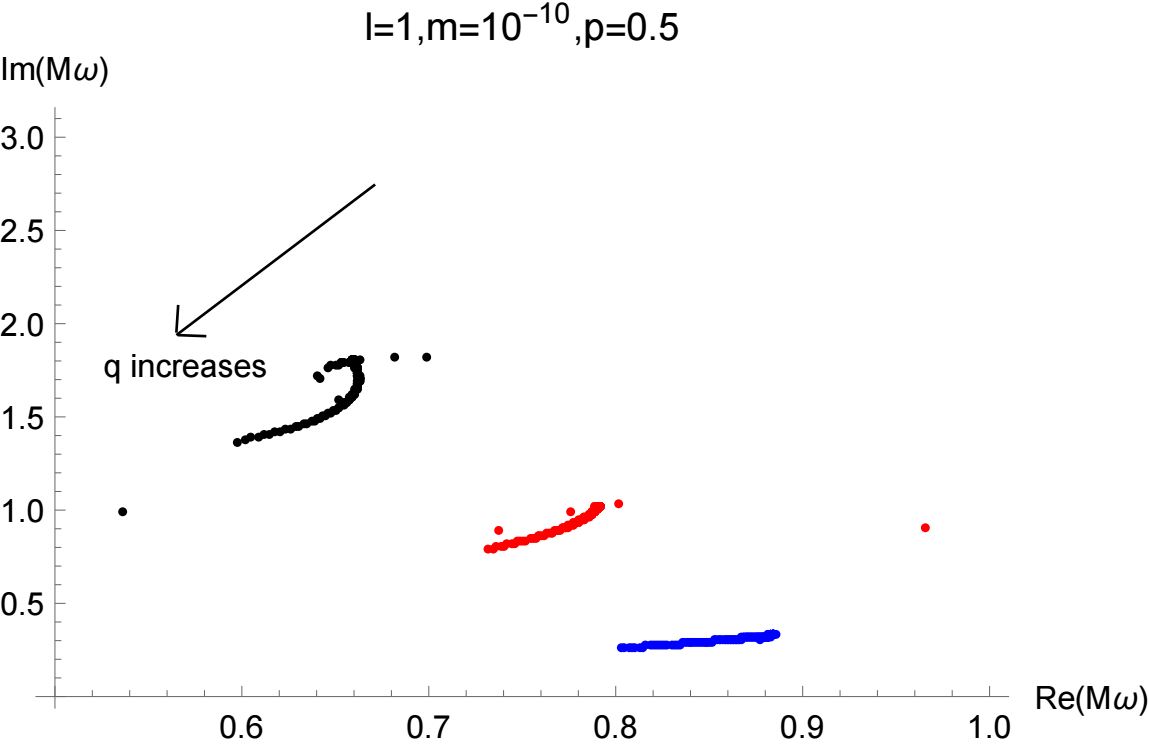}
        \caption{The migration of the first 3 or 4 overtones under the parameter $m,l$ fixed and simply changing $p$ from 0 to 1 (left) or simply changing $q$ from 0 to 0.8 (right). Here and in the following calculation $q$ does not range from 0 to 1 to avoid being captured in the forbidden region.}
        \label{fig:qi}
\end{figure}

  From the top right and bottom right panel of Fig.\ref{fig:001} one can confirm that when black hole mass $M$ is fixed, the cosmological constant $\Lambda$ mainly relies on $p$, while $\alpha$ mainly relies on $q$. For specific perturbations we firstly study the corrections caused by quantum effect itself to the original Schiwarzschild-dS QNMs, under different overtone number $n$ and angular momentum number $l$. So we change parameter $q$ under a constant parameter $p$, and when $q$ reduces to 0 it's just the case for Schiwarzschild-dS spacetime. Also, one can study the perturbation caused by cosmological constant: when $q$ is fixed and $p$ changes from 0 to 1, and the results are shown in Fig.\ref{fig:qi}. 
  
  The general rules are as follows: the QNMs of the Proca field are more unstable under parametric perturbations caused by $p$ than by $q$. It's rather easy to understand the reason: the change in $p$ finally ends up in a quite different result in the asymptotic behavior of the Proca field: when $p$=0 the Proca behaves as the really massive field as in the flat spacetime. For $p$ or $q$, when they approach 1 together, the three horizons merge and the effective potential diminishes, so the QNMs reduces to near 0 and is reflected by Fig.\ref{fig:qi}. In the migrating process there are several points for these first overtones to behave Type I instability where there are obvious discontinuity on some of these curves, and this is more likely to happen for higher overtones. By comparing the shapes of the migration curves in Fig.\ref{fig:qi} with some previous works of the pseudospectrum of QNMs~\cite{Cao:2024oud}, it's very clear that even for the most stable fundamental modes, they still migrate in the unrestricted lines. And this originates from the fact that the operator represented by the discontinuous matrix $L$ is non self-adjoint, and there is no guarantee of the stability of its eigenvalues under perturbation.
 
\begin{figure}[htbp]
\includegraphics[width=0.43\textwidth]{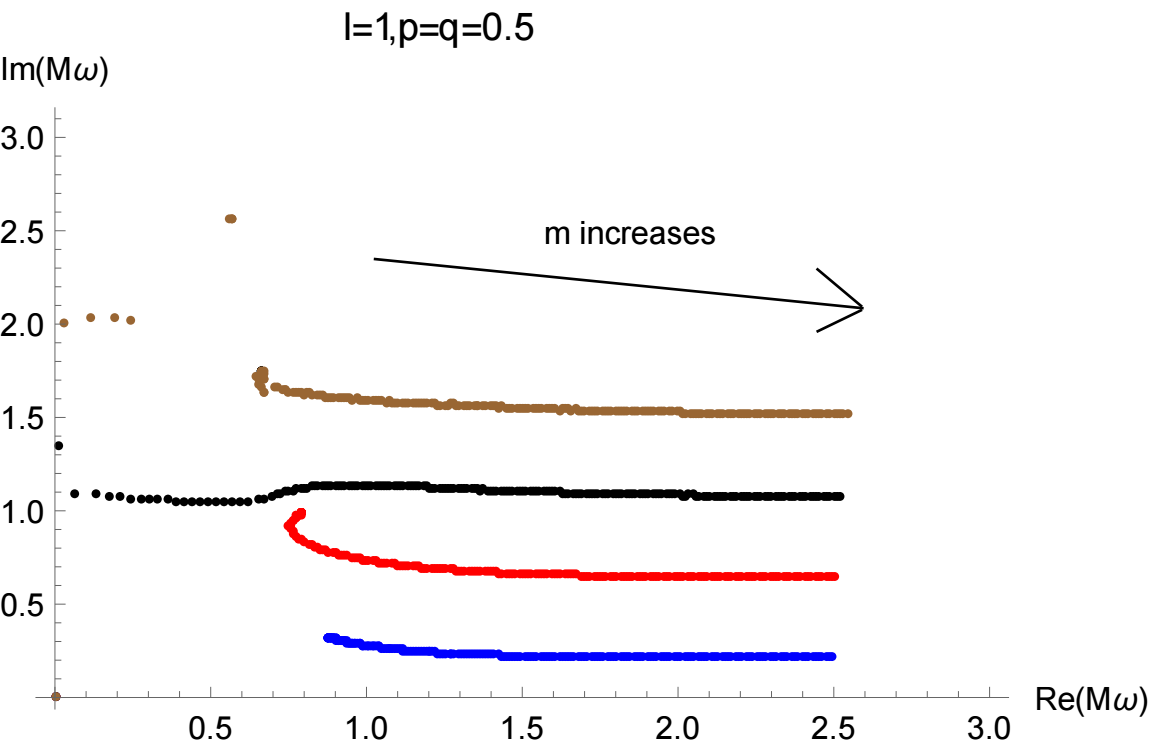}
\includegraphics[width=0.43\textwidth]{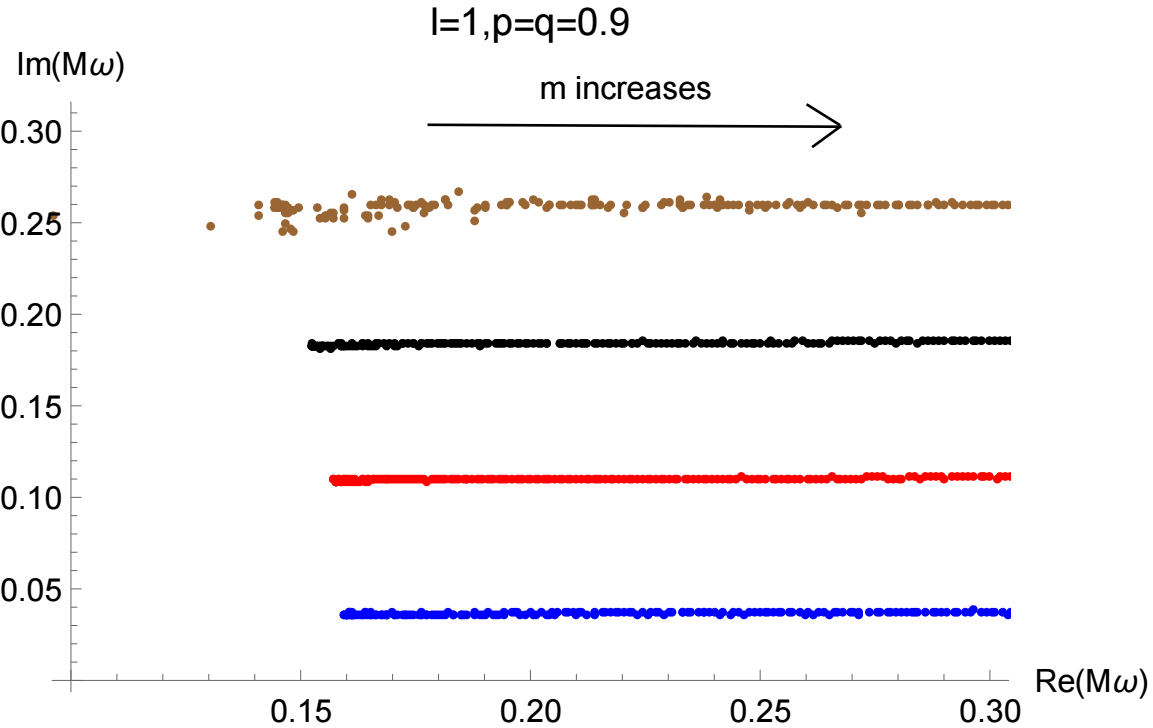}
        \caption{The migration of the first 3 or 4 overtones under the parameter $l,p,q$ fixed and simply changing $m$ from 0 to 1.}
        \label{fig:qs}
\end{figure}

Another aspect to study is the mass constant $m$, which, according to the expression of the effective potential, can simply reduce the possible difference caused by different $l$ up to a common ratio for all QNMs, so we don't need to change $l$ any longer. In Fig.\ref{fig:qs}, the QNMs migration of first overtones under the change in $m$ is shown, and it is clear that $m$ mostly contributes to a increase in the real part of the QNMs but does nothing to the imaginary part, so the increase of the oscillating rate is expected under the growth in the Proca mass, but no change in the decay rate. The rule for massive fields, the first overtone of the QNMs becoming very small in its imaginary part as $m$ increases, which, according to a lot of previous researches, are called quasi-resonance, is no longer observed. We find $m$ to have (at least for real part) influence nearly opposite from $p$ and $q$, which warns that there may be a possible ``mass concealment" effect. 

\begin{figure}[htbp]
\includegraphics[width=0.43\textwidth]{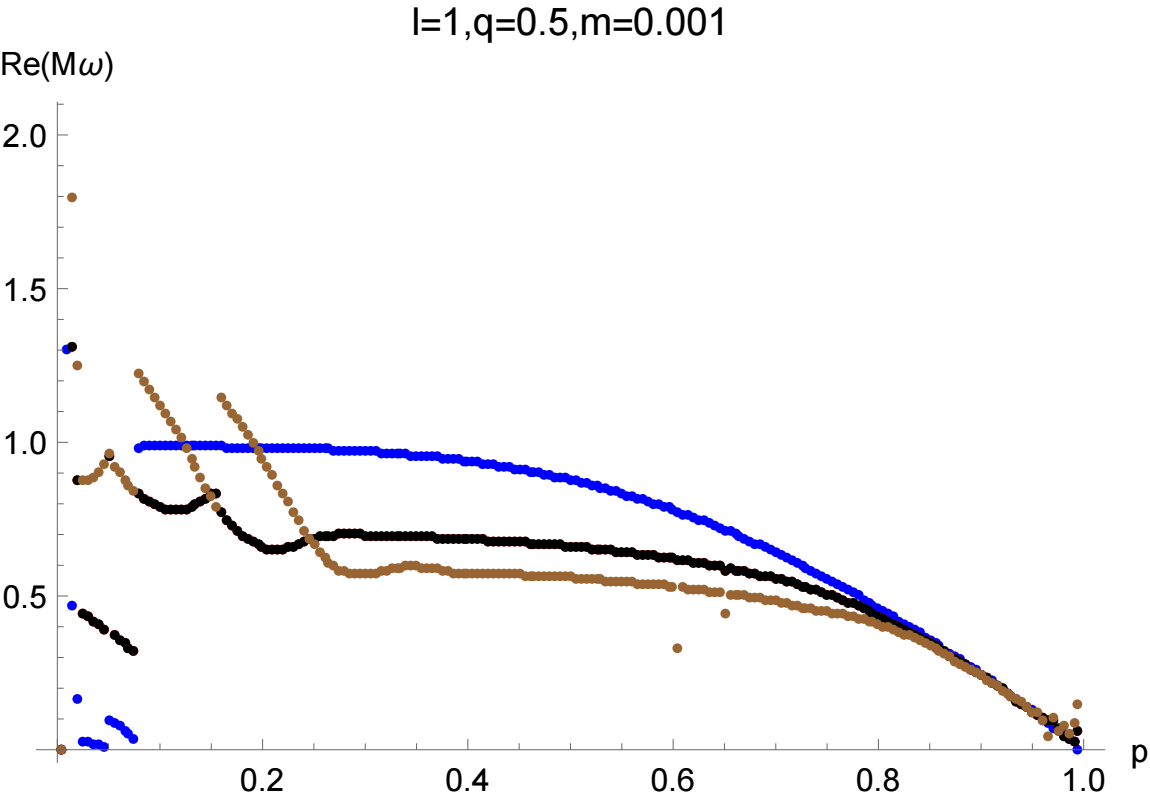}
\includegraphics[width=0.43\textwidth]{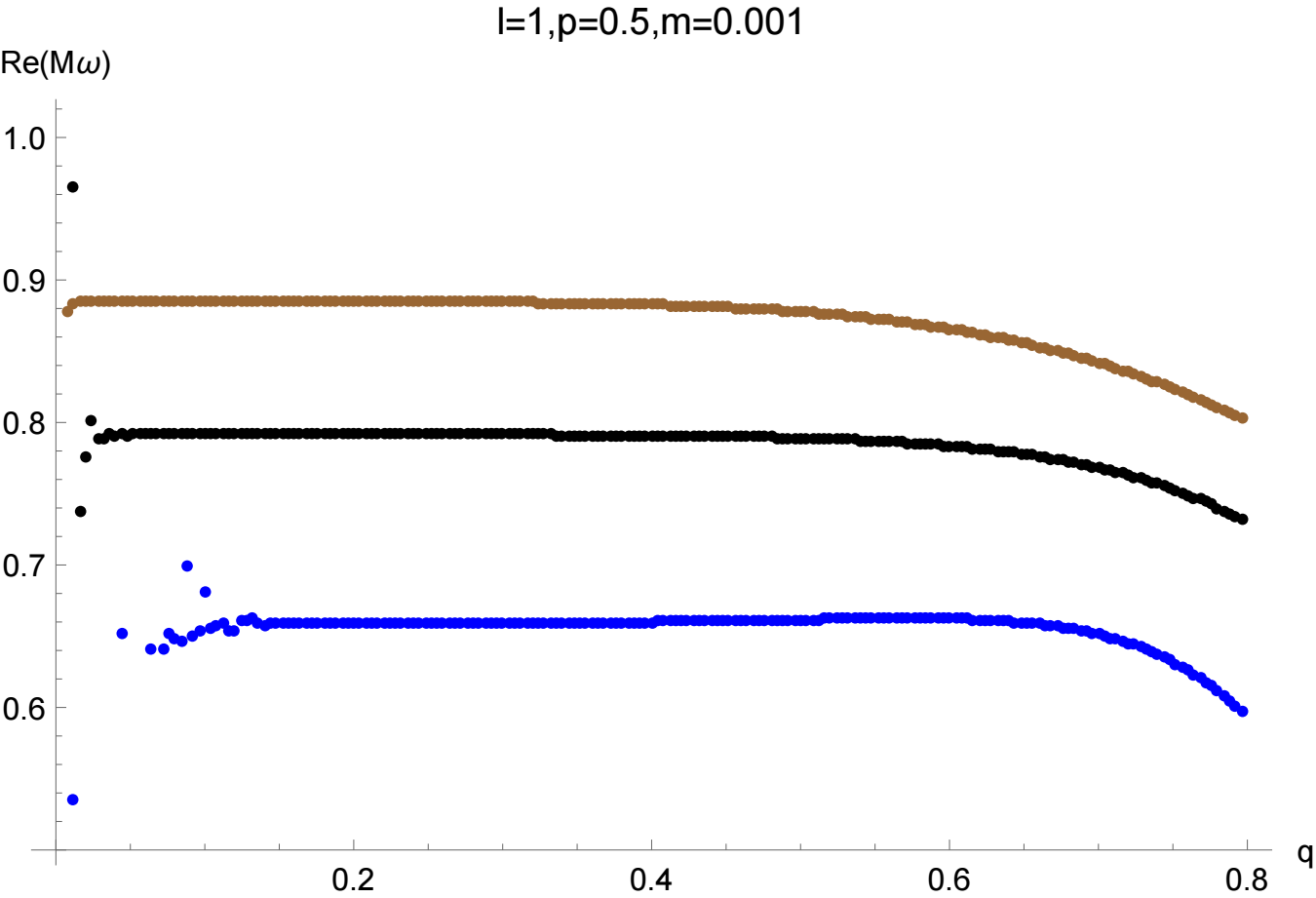}
        \caption{The change of the real part of the first 3 overtones under the parameter $l,m$ fixed and simply changing $p$ from 0 to 1 (left) or simply changing $q$ from 0 to 0.8(right).}
        \label{fig:q}
\end{figure}

\begin{figure}[htbp]
\includegraphics[width=0.43\textwidth]{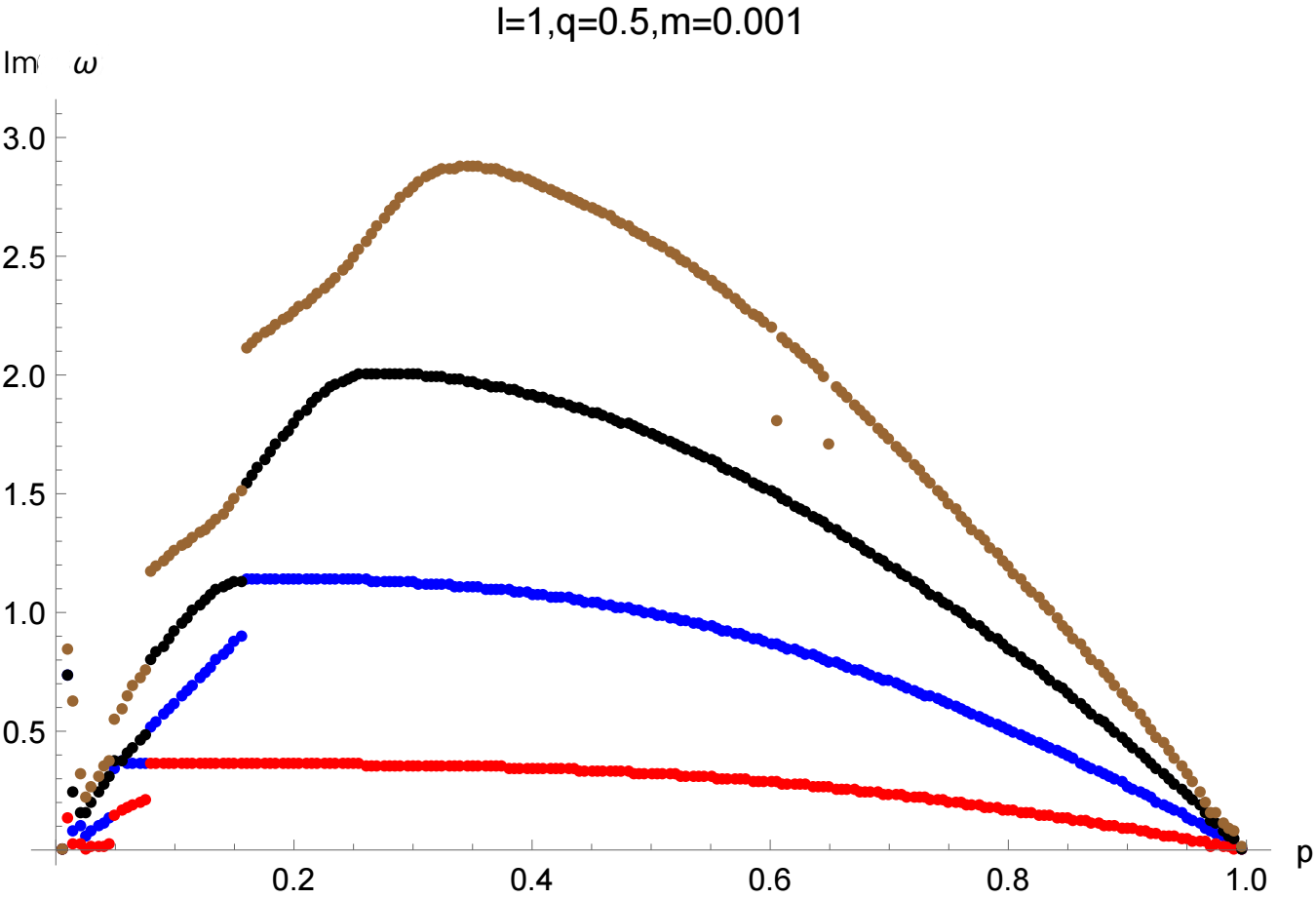}
\includegraphics[width=0.43\textwidth]{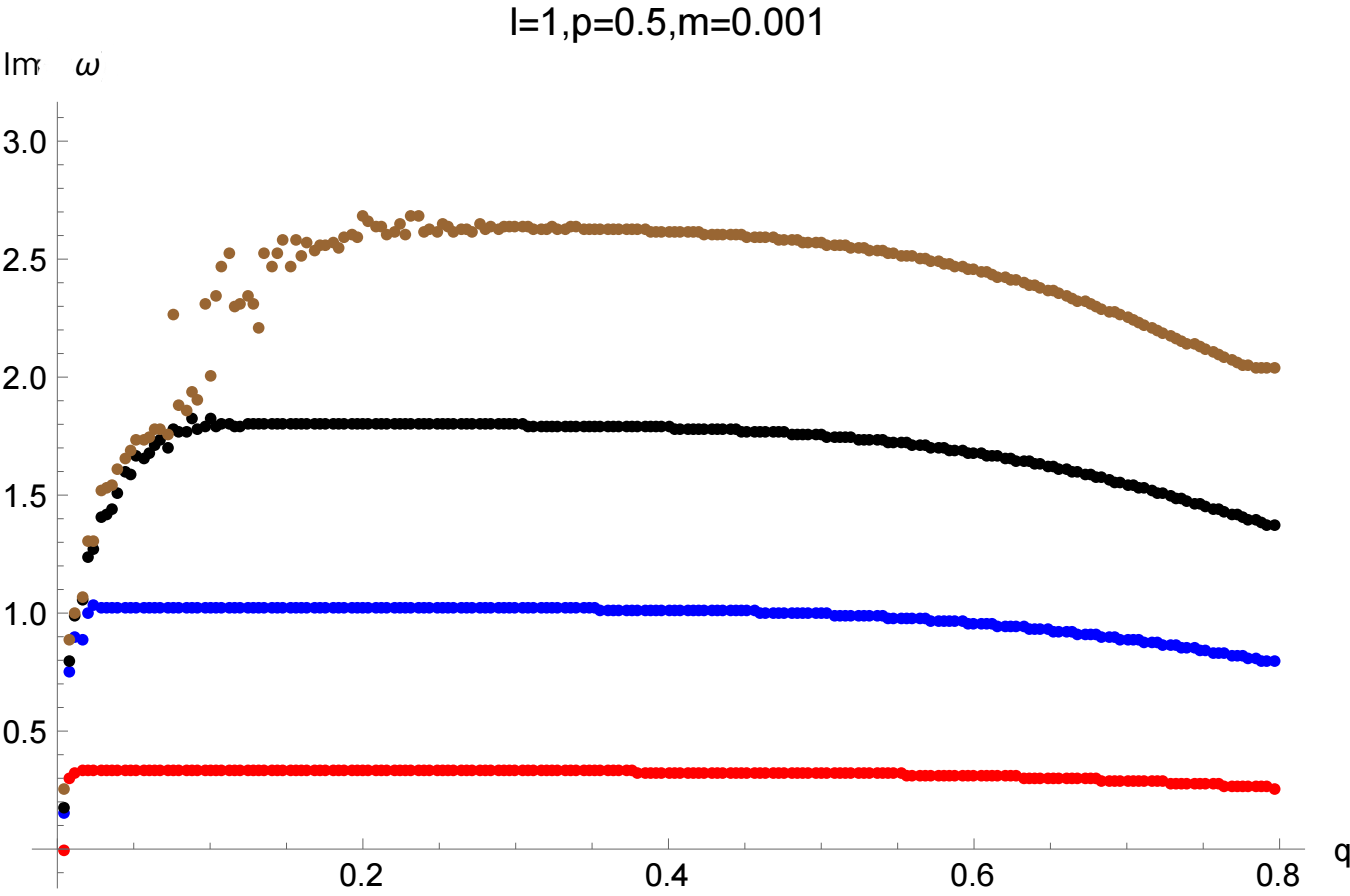}
        \caption{The change of the imaginary part of the first 4 overtones under the parameter $l,m$ fixed and simply changing $p$ from 0 to 1 (left) or simply changing $q$ from 0 to 0.8(right).}
        \label{fig:qqq}
\end{figure}

After attaining the general migration of the QNMs, it's meaningful to study the specific migration of the real parts and imaginary part of QNMs separately. Some results are shown in Fig.\ref{fig:q} and Fig.\ref{fig:qqq}. In the first figure of Fig.\ref{fig:qqq}, a phenomenon of overtone replacement is observed, as when $p$ approaches 0, Type I instability occurs and continuous curves of overtone migration are disrupted by suddenly occurring new branches.
This phenomenon rapidly lowers the total imaginary part of QNMs, with greater possibility to move to quasi-resonance, which indicates the fundamental QNM ending up in imaginary part very close to 0. This, together with the result of changing $m$ but keeping $p$ a constant, once again illustrates the asymptotic behavior of the field is more important than the mass itself.

\subsection{Pseudospectrum }\label{sec:42}

Given $\epsilon>0$, the $\epsilon$-pseudospectrum $\sigma_{\epsilon}(A)$ of an operator $A$ is defined as~\cite{Cao:2024oud}
\begin{eqnarray}\label{pseudospectrum_definition}
	\sigma_{\epsilon}(A)&=&\{z\in\mathbb{C}:\lVert R_{A}(z)\rVert=\lVert(z\mathbb{I}-A)^{-1}\rVert>1/\epsilon\}\, ,
\end{eqnarray}
where $R_{A}(z)$ is called the resolvent operator. Although there are many other equivalent definitions, we can choose this as the most suitable one for visualizing the pseudospectrum. It is clear that $\epsilon\to0$ corresponds to the set $\sigma_\epsilon(A)\rightarrow\sigma(A)$, whose elements are nothing but the spectrum $\omega_n$. The quantity $\epsilon$ serves as a bridge between points in $\sigma_\epsilon(A)$ and the spectrum $\omega_n$, and in physical circumstance it offers a clear interpretation of perturbations. One can easily read out quantify the spectrum (in)stability of the operator $A$ from the shape and size of the $\epsilon$-pseudospectrum regions. A usual deduce principle is, when the contour lines of the pseudospectrum forms a concentric circle of the spectrum, the operator considered is spectrally stable, otherwise it is spectrally unstable. With large enough perturbations, the spectrum will become unstable due to the black hole boundary condition which leads to dissipation~\cite{PhysRevD.104.084091,PhysRevD.108.104002}.

About addressing the construction of pseudospectrum in
a numerical approach, we can use Chebyshev differentiation matrices and Chebyshev-Lobatto $N$-point grids to produce $L$ matrix approximates. Once the operator is discretized,
the construction of the pseudospectrum requires the evaluation of matrix norms. A standard practical choice 
involves the matrix norm induced from the 2-norm in the vector space, which leads to the following
rewriting~\cite{PhysRevX.11.031003}:
\begin{eqnarray}
    \sigma^{\epsilon}_{2}(\textbf{L})&=&\{z\in\mathbb{C}:\sigma^{min}(z\mathbb{I}-\textbf{L})<\epsilon\}\,,
\end{eqnarray}
where $\sigma^{min}(\textbf{M})$ stands for the smallest singular value of the matrix $\textbf{M}$. The most significant advantage to choose the 2-norm is the convenience and exemption from constructing ``energy norm", which is easy to loose effectiveness for systems with more than 1 degree of freedom, or the energy is not positive definite, such as in Kerr spacetime.

\begin{figure}[htbp]
	\centering
\includegraphics[width=0.4\textwidth]{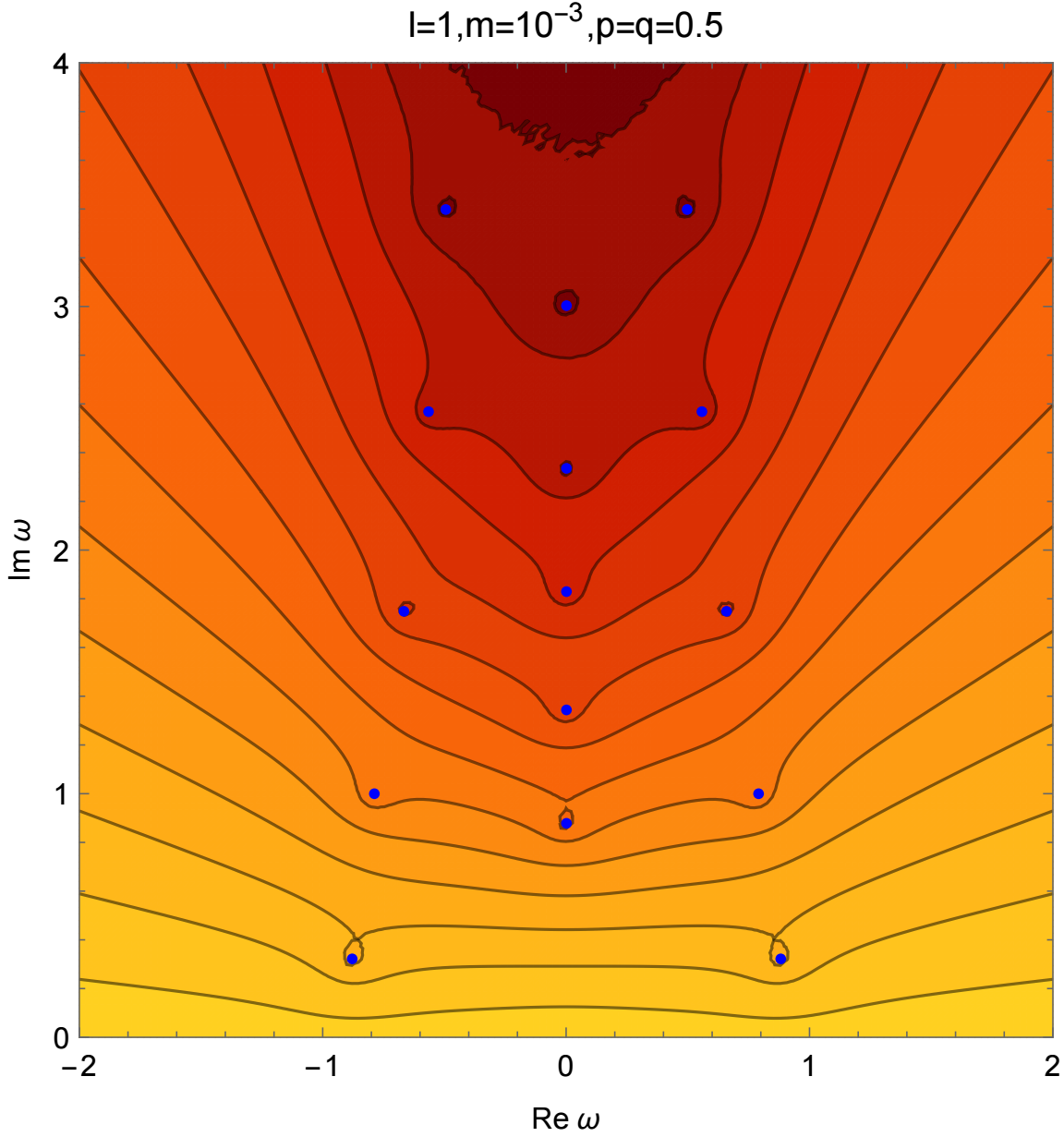}
\includegraphics[width=0.06\textwidth]{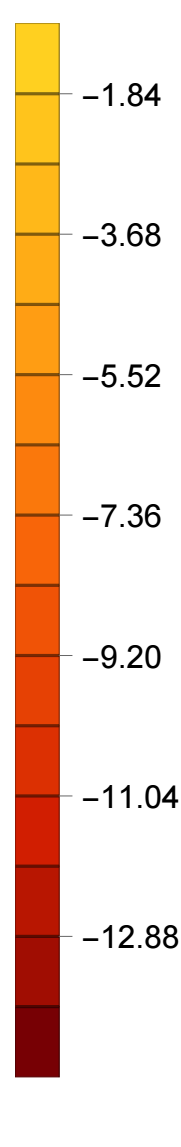}
\includegraphics[width=0.41\textwidth]{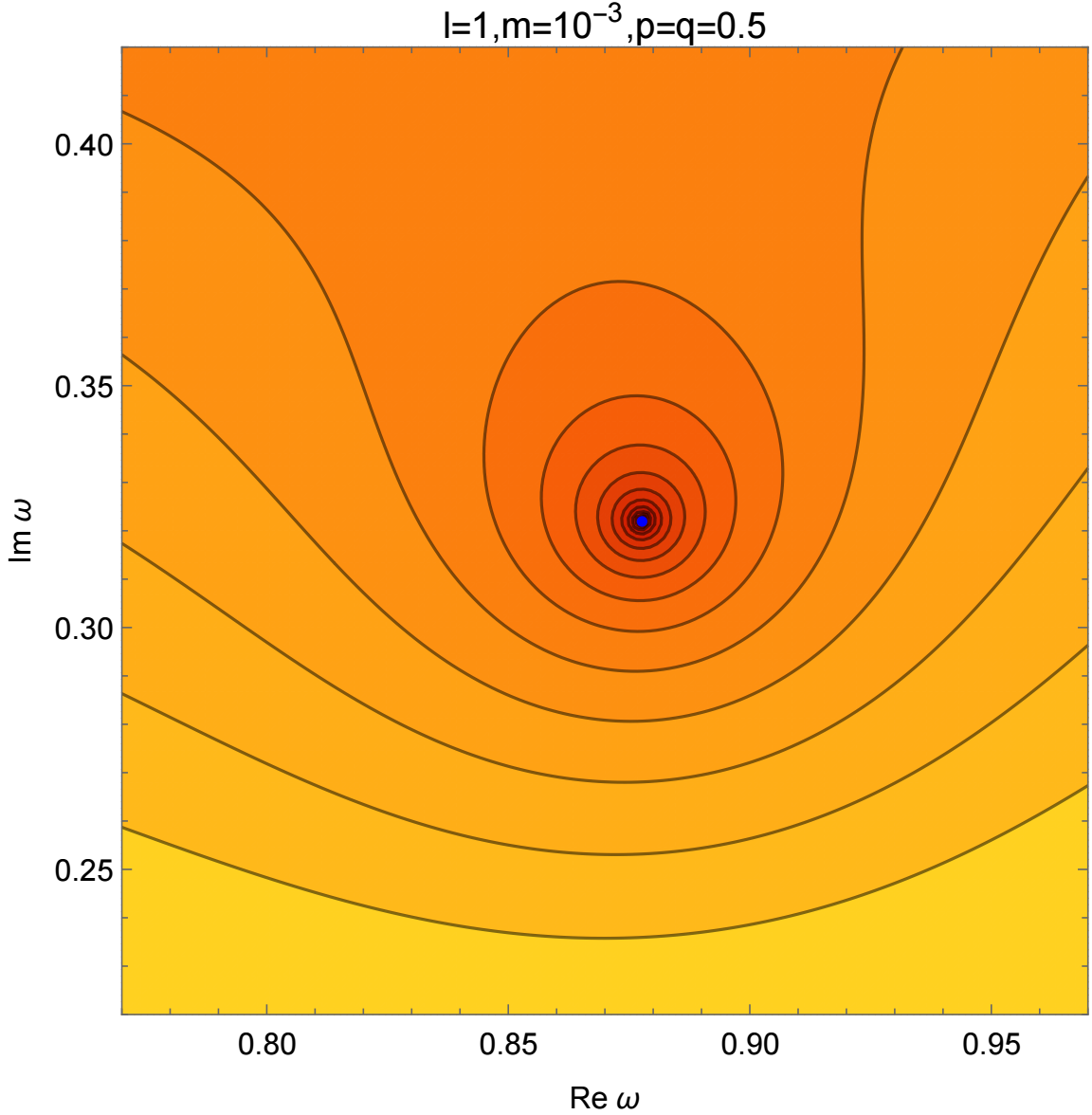}
\includegraphics[width=0.05\textwidth]{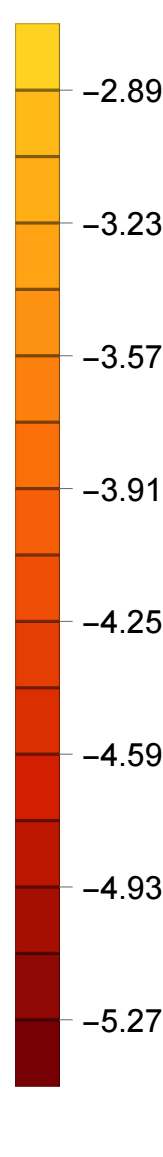}

        \caption{The $\epsilon$-pseudospectrum of the quantum corrected black hole with $q=p=0.5,m=10^{-3}$ and $l=1$. The solid contour lines correspond to various choices of $\log_{10}\epsilon$. The left panel represents the zoomed-out view of the $\epsilon$-pseudospectrum. The right panel represents the zoomed-in view of the $\epsilon$-pseudospectrum around the fundamental mode $n=0$. }
        \label{fig:p1}
\end{figure}

\begin{figure}[htbp]
	\centering
\includegraphics[width=0.4\textwidth]{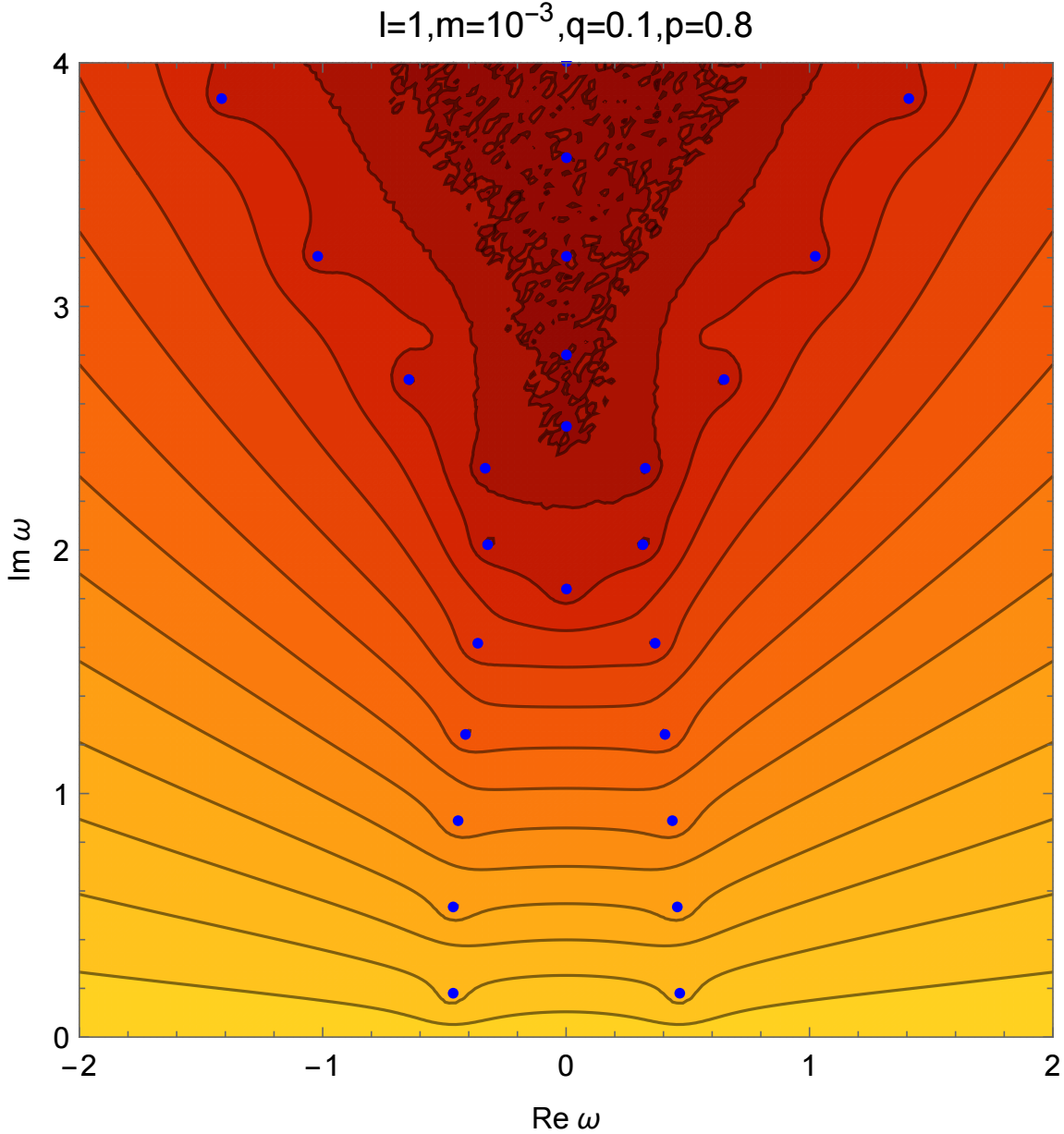}
\includegraphics[width=0.045\textwidth]{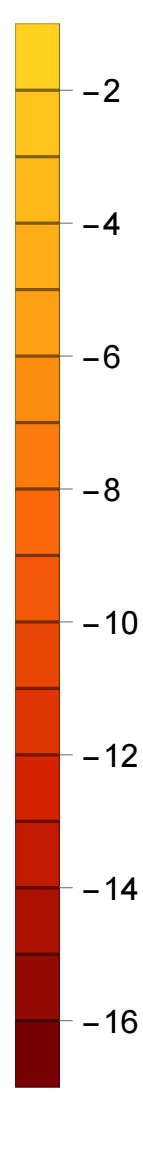}
\includegraphics[width=0.41\textwidth]{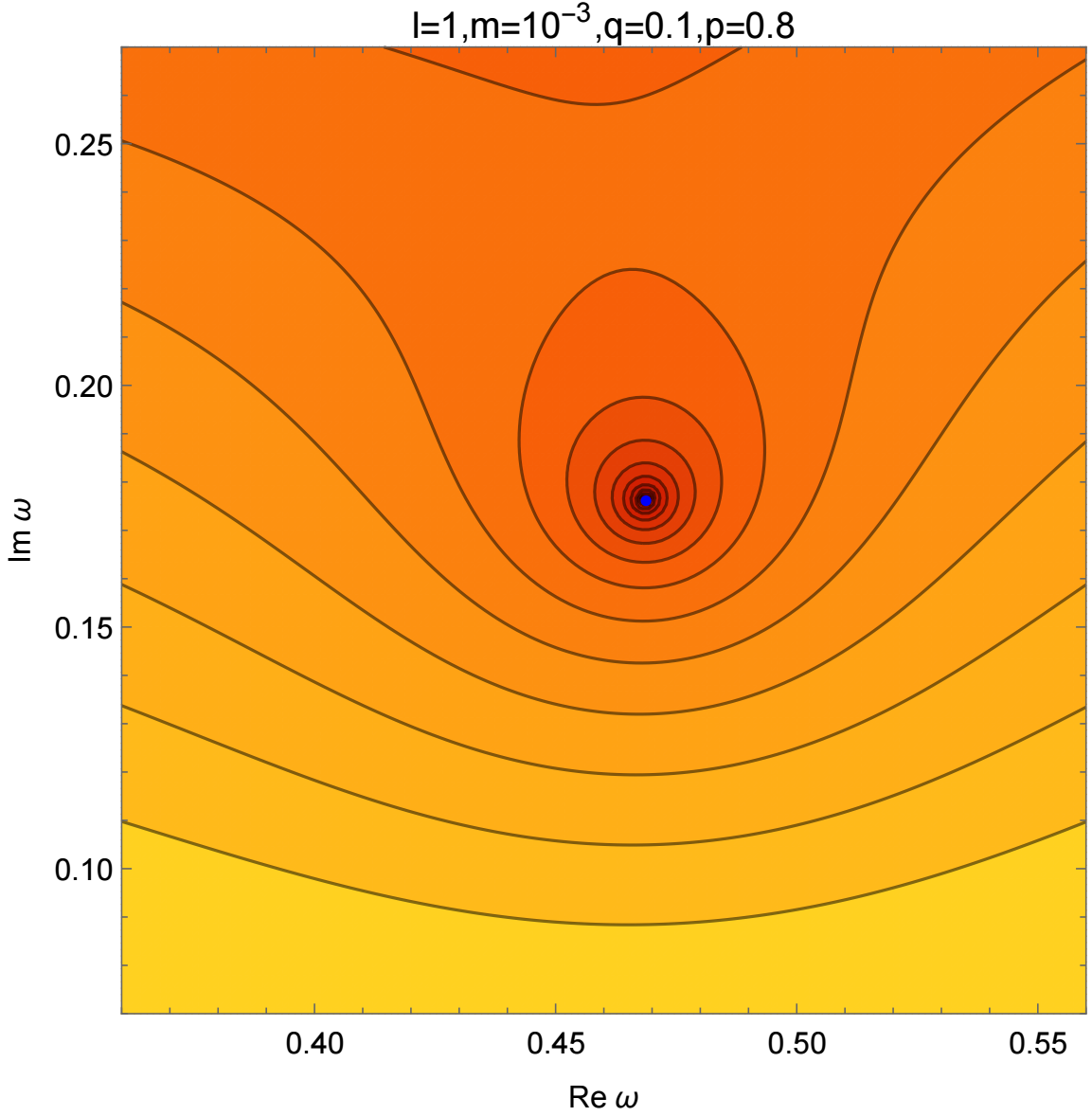}
\includegraphics[width=0.05\textwidth]{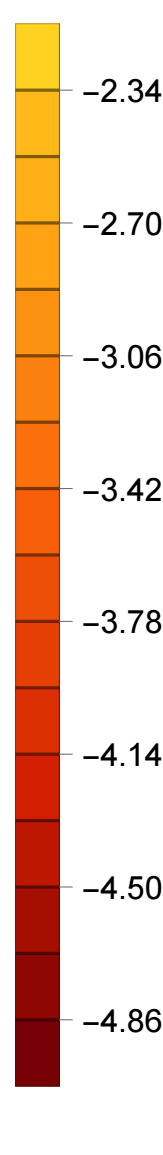}

        \caption{The $\epsilon$-pseudospectrum of the quantum corrected black hole with $q=0.1,p=0.8,m=10^{-3}$ and $l=1$. The solid contour lines correspond to various choices of $\log_{10}\epsilon$. The left panel represents the zoomed-out view of the $\epsilon$-pseudospectrum. The right panel represents the zoomed-in view of the $\epsilon$-pseudospectrum around the fundamental mode $n=0$.}
        \label{fig:p2}
\end{figure}

In Fig.\ref{fig:p1} and Fig.\ref{fig:p2}, the contour map of the pseudospectrum is shown, and each contour line corresponds to a constant $\sigma^{min}(z\mathbb{I}-\textbf{L})$ on the $z$ plane. As the imaginary part increases, the instability of QNMs get more and more severe. The non-Hermitian of the operator $L$ is reflected by those unclosed contour lines, although from the right panel of Fig.\ref{fig:p1} one can still see very tiny enclosed circles around a QNM. These circles offer a important index of whether a numerical result of a QNM is believable: if the calculation error is found to be even larger than the outermost circle's $\epsilon$ value, than the result would be non-convergent. Also, the contour lines show pretty good accordance to the unclosed migration flow in Sec.\ref{sec:10}, but here we can even extend the meaning of a perturbation to a arbitrary shape, given a certain amplitude. We can therefore conclude that the results shown in Fig.\ref{fig:qi} is not coincident.

Now that we have got the characteristics of the perturbation of QNMs in the frequency domain, we naturally want to get more knowledge about the evolution in the time domain as well, which will be illustrated in the next section. 

\section{evolution in the time domain}\label{sec:4}

We first need to get the expression of the Green function in the frequency domain. By introducing $s=i\omega$, and remaining the definition of coefficients for a specific $\omega$ in Eq.(\ref{e1}) and Eq.(\ref{f1}), we have 
\begin{eqnarray}
    G(s,x,x')=\frac{1}{2sa(s)}(\theta(x-x')\phi_{s}(x')\psi_{s}(x)+\theta(x'-x)\phi_{s}(x)\psi_{s}(x'))\,,
\end{eqnarray}
and then in time domain we can use the inverse Laplace transformation:
\begin{eqnarray}\label{ld}
    G(t,x',x)=\frac{1}{2\pi i}\int_{s_{0}-i\infty}^{s_{0}+i\infty}\mathrm{d}s\mathrm{e}^{st}G(s,x,x')\,,
\end{eqnarray}
where $s_{0}$ is a positive number. Through residue theorem we can transfer this to three parts, so in general, as illustrated in~\cite{yang2024spectralinstabilityblackholes}, there should be three main stages of the evolution: 1. Precursor, mainly coming from integration along the large semicircle on the half plane with $|s|\rightarrow \infty$. 2. Ringdown, mainly coming from the poles of $a(s)$, which is related to QNMs, (actually related to greybody factor which is stable under perturbation~\cite{rosato2024ringdownstabilitygreybodyfactors}). 3. Power-law tail, mainly coming from branch cut on the negative real axis (NRA). 

However, as in the qOS-dS spacetime, the effective potential decays exponentially, there is actually no branch cut on the NRA. So there is no third stage at all, which means no power-law tail is expected. For convenience  we focus on the case when the initial data mainly lies on positions far in the future of the peak of the potential. Actually later numerical calculation shows only little difference for the different observer's position in the main part of time, so we suppose the observer being in infinite future as well, but on the left side of the source. Now we only need to consider the asymptotic behavior of the Green function $G(t,x',x)$ as $x'\rightarrow \infty$, $x\rightarrow\infty$ and $x<x'$, where $x'$ is the position of the source and $x$ is the position of the observer. So now through the definition in Eq.(\ref{e1}) we have 
\begin{eqnarray}
    G(s,x',x)=\frac{1}{2s}\Big{[}\psi_{-s}(x)\psi_{s}(x')+\frac{b(s)}{a(s)}\psi_{s}(x)\psi_{s}(x')\Big{]}\,,\qquad x,x'\rightarrow\infty\,,
\end{eqnarray}
so we naturally divide Eq.(\ref{ld}) into two parts:
\begin{eqnarray}
    G^{(1)}(t,x',x)=\mathcal{L}^{-1}\Big{[}\frac{1}{2s}\psi_{-s}(x)\psi_{s}(x')\Big{]}\,, \qquad
    G^{(2)}(t,x',x)=\mathcal{L}^{-1}\Big{[}\frac{b(s)}{2sa(s)}\psi_{s}(x)\psi_{s}(x')\Big{]}\,,
\end{eqnarray}

For high frequency limit $|s|\rightarrow\infty$ one can suppose~\cite{PhysRevD.86.024021}
\begin{eqnarray}\label{iu}
   \psi_{s}(x)=\mathrm{e}^{-sx}\Big{[}1-\frac{g_{1}(x)}{s}+\frac{g_{2}(x)}{s^2}\Big{]}+O(\frac{1}{s^3})\,,\qquad x\rightarrow\infty\,,
\end{eqnarray}
and by inserting this into the wave equation we can solve $g_{1}(x)$ and $g_{2}(x)$ order-by-order in $s$:
\begin{eqnarray}
    g_{1}(x(r))=\frac{1}{2}\Big{[}\frac{l(l+1)}{rr_{0}}+m^2\Big{]}(r-r_{0})\,,
\end{eqnarray}
\begin{eqnarray}
    g_{2}(x(r))=\frac{1}{8}(r-r_{0})^2\Big{[}\frac{l(l+1)}{r_{0}r}+m^2\Big{]}^2+\frac{1}{4}\Big{[}\frac{l(l+1)}{r^2}+m^2\Big{]}f(r)\,,
\end{eqnarray}
in which we have guarantee that as $r\rightarrow r_{0}$ both $g_{1}(x)$ and $g_{2}(x)$ reduce to 0. Simply replacing $s$ by $-s$ and $r_{0}$ with $r_{+}$ we get the expression for $\phi_{s}(x)$ as $x\rightarrow-\infty$. To get the expression for $a(s)$ we need to approximate when $x\approx 0$
 \begin{eqnarray}
     2sa(s)\approx \psi_{s}(0)\frac{\mathrm{d}\phi_{s}(0)}{\mathrm{d}{x}}-\phi_{s}(0)\frac{\mathrm{d}\psi_{s}(0)}{\mathrm{d}x}=2s+(r_{0}-r_{+})\Big{[}\frac{l(l+1)}{r_{+}r_{0}}+m^2\Big{]}+O(\frac{1}{s})\,,
 \end{eqnarray}
\begin{eqnarray}
    2sb(s)\approx \phi_{s}(0)\frac{\mathrm{d}\psi_{-s}(0)}{\mathrm{d}{x}}-\psi_{-s}(0)\frac{\mathrm{d}\phi_{s}(0)}{\mathrm{d}x}=\frac{r_{0}-r_{+}}{2s}\Big{[}\frac{l(l+1)}{r_{+}r_{0}}+m^2\Big{]}+O(\frac{1}{s^2})\,,
\end{eqnarray} 
\begin{eqnarray}
  G^{(1)}(s,x',x)\approx\frac{\mathrm{e}^{s(x-x')}}{2s} \,,\qquad G^{(2)}(s,x',x)\approx\frac{(r_{0}-r_{+})}{8s^3}\Big{[}\frac{l(l+1)}{r_{+}r_{0}}+m^2\Big{]}\mathrm{e}^{-s(x+x')}\,, 
  \end{eqnarray}
and now we divide the whole time domain into three parts. First of all, when $t<x'-x$, then for both $G^{(1)}(t,x',x)$ and $G^{(2)}(t,x',x)$, we need to choose the large semicircle on the right-half plane so both semicircle contour integrals are 0 according to Jordan Theorem, with no residue in their contours as well. Now $G(t,x',x)=0$. This result is natural considering the not-so-good geometric optical approximation made in Eq.(\ref{iu}) when $x'$ is still finite. 

\begin{figure}[htbp]
	\centering
\includegraphics[width=0.4\textwidth]{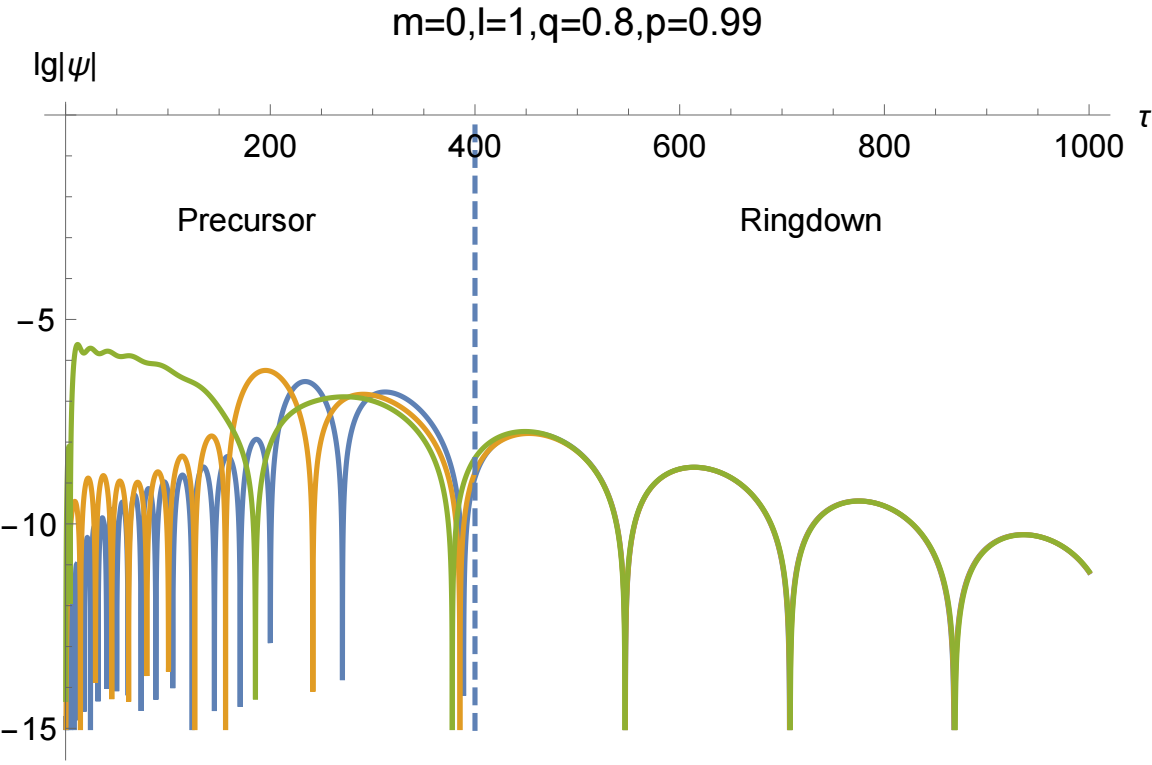}
\includegraphics[width=0.07\textwidth]{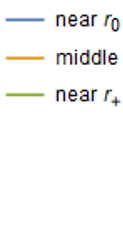}
\includegraphics[width=0.4\textwidth]{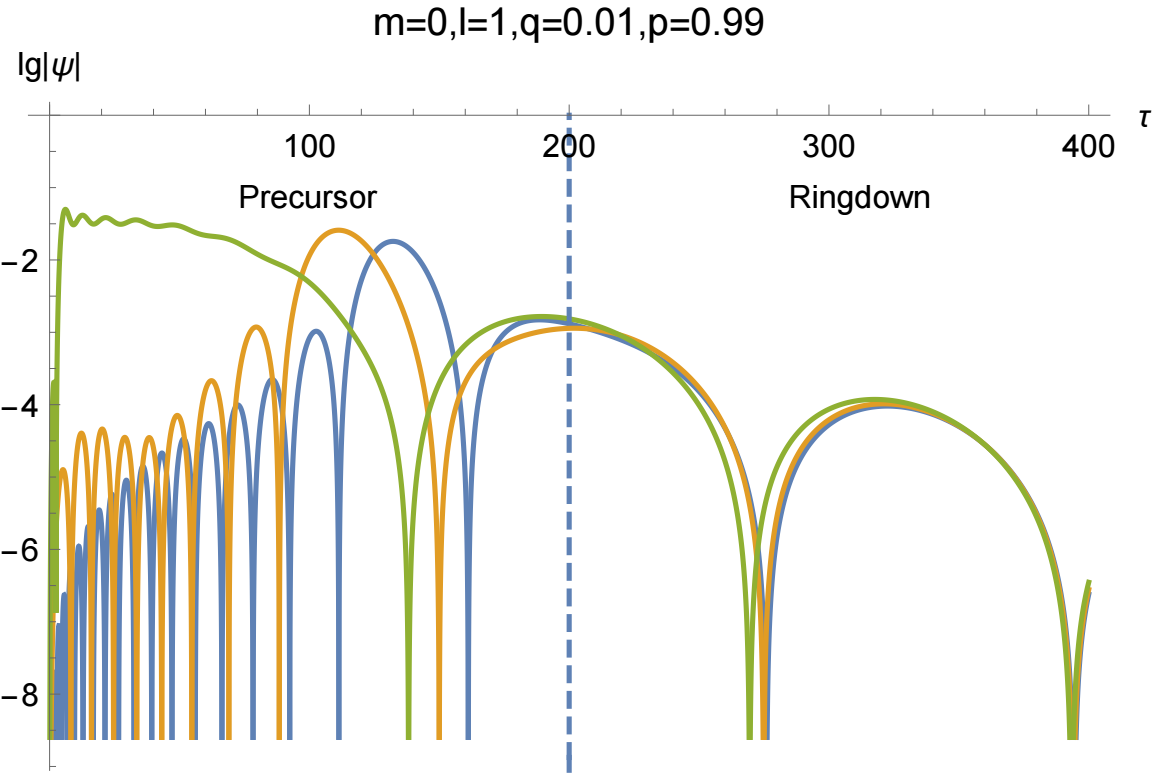}
    
        \caption{The time evolution of the Proca field on different positions under the parameter $m=0,l=1,p=0.99$ and different $q$, with the initial data set in (\ref{cm}) and (\ref{cmb}), in which $x_{0}=r_{*}((10r_{0}+r_{+})/11)\gg 0$. }
        \label{fig:ww}
\end{figure}

When $x'-x<t<x'+x$, $G^{(2)}(t,x',x)$ is still 0 while for $G^{(1)}(t,x',x)$ we have to choose the semicircle on the left-half plane, and now the total contribution is 
\begin{eqnarray}
   G^{(1)}(t,x',x)=\mathrm{Res}\Big{[}G^{(1)}(s,x,x'),0\Big{]}\,,
\end{eqnarray}
the wave form nearly does not change with time, and it is exactly the Precursor. However, precisely  the contribution from both the contour integral and $G^{(2)}(s,x',x)$ are nonzero, and this can be proven through numerical results (e.g.Fig.\ref{fig:ww}), 
in which waveform slightly changes and even grows with time in Precursor.

When $t>x'+x$, we arrive at the Ringdown stage, during which we have to choose the semicircle on the left-half plane for both terms, and now the result is
\begin{eqnarray}
G(t,x',x)=\mathrm{Res}\Big{[}G(s,x,x'),0\Big{]}+\sum_{s_{n}}\frac{\mathrm{e}^{s_{n}t}b(s_{n})\psi_{s_{n}}(x)\psi_{s_{n}}(x')}{2s_{n}a'(s_{n})}+\ldots\,,
\end{eqnarray}
where $s_{n}=i\omega_{n}$ and $\omega_{n}$ is a certain QNM whose imaginary part is always positive, which means it contains contributions of all the QNMs. The ellipsis stands for those QNMs corresponding to higher-order poles or equivalently, Type-I instability. Now it's very clear that the evolution in this stage is exponential decay. 

\begin{figure}[htbp]
	\centering
\includegraphics[width=0.4\textwidth]{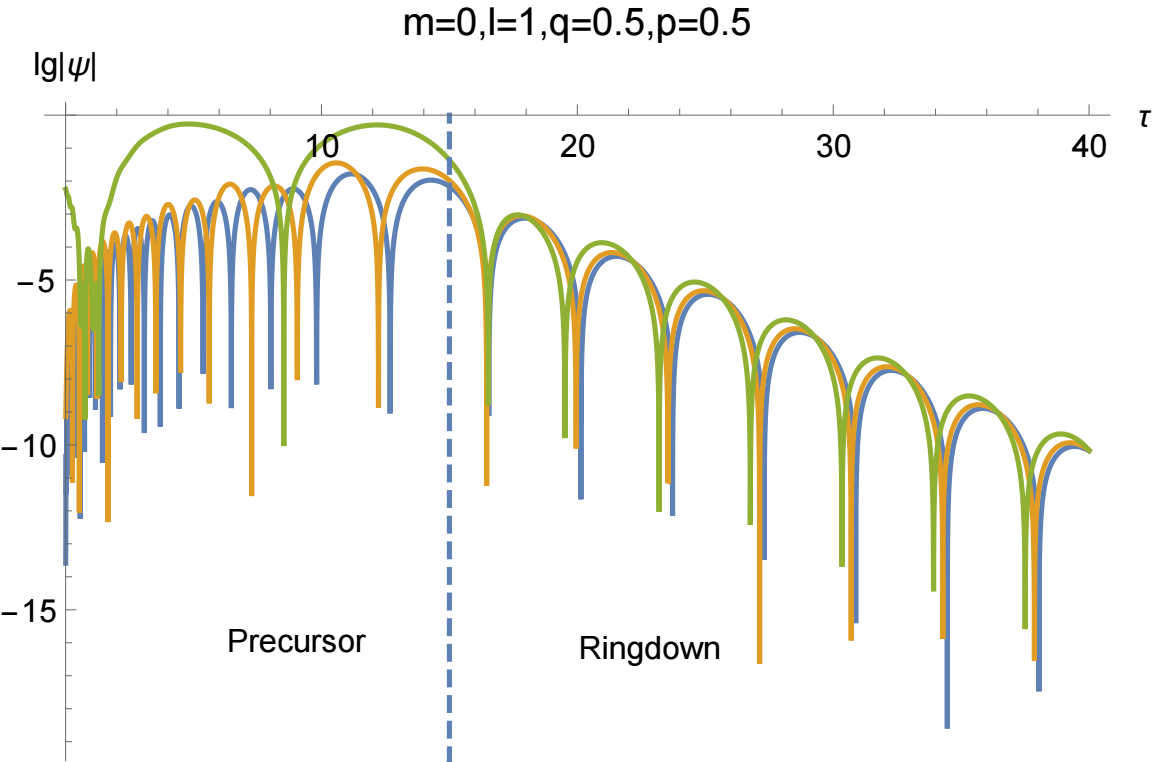}
\includegraphics[width=0.07\textwidth]{021.png}
\includegraphics[width=0.4\textwidth]{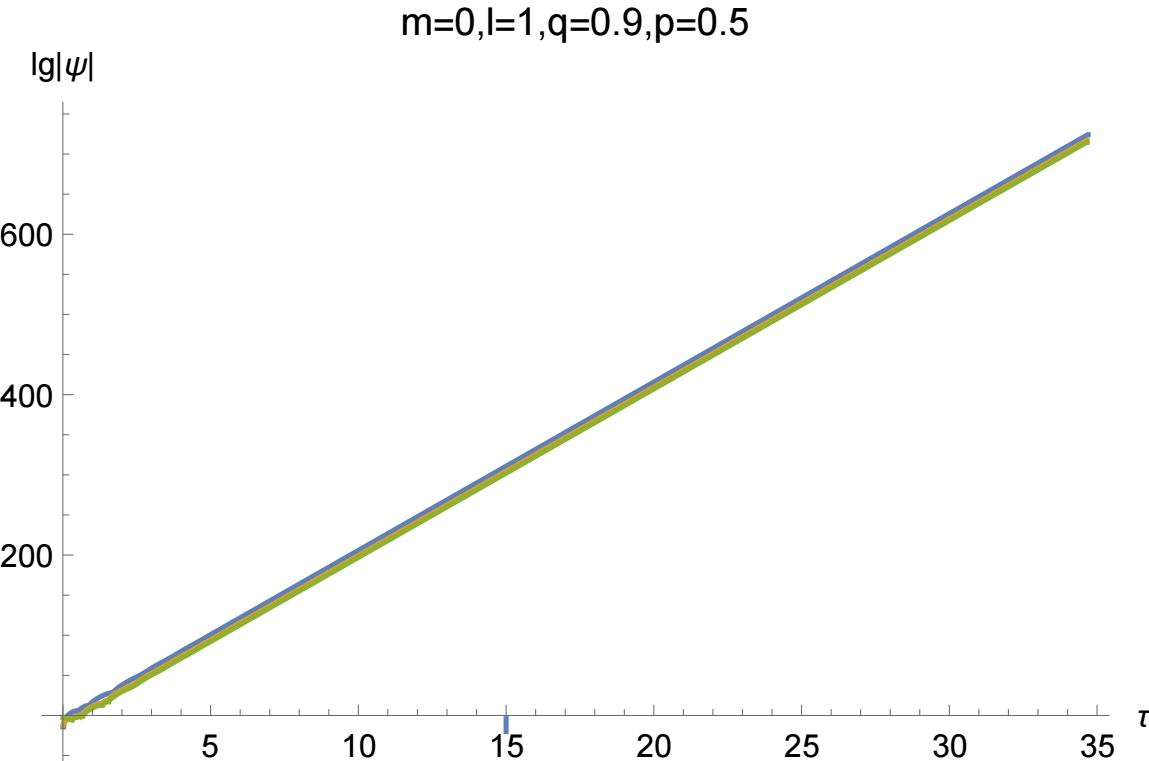}
    
        \caption{The time evolution of the Proca field on different positions under the parameter $m=0,l=1,q=0.5$ and different $p$, with the initial data set in (\ref{cm}) and (\ref{cmb}), in which $x_{0}=r_{*}((100r_{0}+r_{+})/101)\gg 0$. }
        \label{fig:www}
\end{figure}

For numerical calculation we follow the methods introduced in~\cite{OBoyle:2022yhp}, that is, on the basis of the already attained $\textbf{L}$ in (\ref{12345}), we can use the Hermite evolution matrix 
\begin{eqnarray}
    \textbf{A}^{(4)}=\Big{(}\textbf{I}-\frac{\Delta \tau}{2}\textbf{L}+\frac{(\Delta \tau)^2}{12}\textbf{L}^2\Big{)}^{-1}\Big{(}\textbf{I}+\frac{\Delta \tau}{2}\textbf{L}+\frac{(\Delta \tau)^2}{12}\textbf{L}^2\Big{)}\,,
\end{eqnarray}
to transform the field strength vector in the way of 
\begin{eqnarray}
    \textbf{u}^{(n+1)}=\textbf{A}^{(4)}\textbf{u}^{(n)}\,,
\end{eqnarray}
in which
\begin{eqnarray}
   \textbf{u}^{(n)}= \begin{bmatrix}
        \Phi(\tau_{0}+n\Delta\tau)\\
        \Psi(\tau_{0}+n\Delta\tau)
    \end{bmatrix}\,.
\end{eqnarray}
Also, we set the similar initial data as in~\cite{PhysRevD.101.104009}:

\begin{eqnarray}\label{cm}
    \Phi(\sigma,\tau=0)=\mathrm{exp}(-\frac{(r_{*}(\sigma)-x_{0})^2}{8r_{+}^2\lambda^2})\,,
\end{eqnarray}
\begin{eqnarray}\label{cmb}
    -\partial_{\tau}\Phi|_{\tau=0}=\partial_{\sigma}\Phi|_{\tau=0}\,,
\end{eqnarray}
here $\lambda=2,x_{0} \gg 0$. Then we have the following results:

\begin{enumerate}

\item When $p$ is larger the decay rate together with the oscillating rate is obviously much slower. Among the two important parameters, only $p$ greatly contributes to the decay rate and the oscillation rate of the field strength, while $q$ contributes just a little. This conclusion can be proven through comparing the two figures in Fig.\ref{fig:ww} and the comparison between the two figures in Fig.\ref{fig:ww} and the first figure in Fig.\ref{fig:www}. This conclusion warns that in real detecting the influence of quantum effect can easily be concealed by non-zero cosmological constant. Still larger $q$ contributes to longer period in Precursor stage and slower decay rate, which is the most practically observable effect.  

\item The so called Forbidden Area mentioned in 
Fig.\ref{fig:fp} once again shows its important meaning in the last picture of Fig.\ref{fig:www}: in this area a non physical exponential explosion is unpreventable. Some modes with their imaginary parts negative is expected under this condition, which again illustrates the hyperboloidal method, at least through the construction introduced in this paper has lost effectiveness. 

\item The amplitude of the Ringdown stage almost has nothing to do with the specific position in $x$ domain. This effect is particularly obvious when $p$ is closer to 1. This shows a tendency towards average distribution of the total energy in  the Precursor stage, and once this process is finished the whole evolution transfers to the Ringdown stage.

\end{enumerate}

\section{conclusions and discussion}\label{sec:5}
In this paper we study the quasinormal modes, their instability and pseudospectrum of the Proca field in quantum corrected qOS-dS spacetime. The qOS-dS solution is a solution under the LQG framework and have different characteristics to traditional RN-dS or Schwarzschild-dS solution. The correction can be summarized as two nonzero parameters, the relative ratio of the inner horizon radius to outer horizon radius $q$ and the relative ratio of the outer horizon radius to cosmological horizon radius $p$. It's a meaningful question of how these two observably practical parameters relies on structural parameters $\alpha$ and $\Lambda$, and through parameterization this question now has an answer. After this was done we need to find the correct moving equation for Proca field in such a spacetime, and the axial perturbation's effective potential is obtained, which can be regarded as a small extension of Regge-Wheeler potential for vector fields. All these are our preparation work (Sec.\ref{sec:1}, Sec.\ref{sec:2}).  

In the first part of our formal work we construct the hyperboloidal framework in the classical way, that is, using the minimal gauge. We examine the effectiveness of this gauge under the most general parameter space and come to find that there really exists a certain parameter region where the spatial characteristic of the $\tau$=constant hypersurface is violated. Following studies (Sec.\ref{sec:4}) further illustrates the importance of guarantee the correctness of the construction. After the hyperboloidal framework and Chebyshev-Gauss-Labatto grid are successfully constructed, we present the QNMs under a certain parameter and certain numerical precision, and then confirm the astringency of the results by increasing the grid number together with increasing the machine precision. 

In the second part we study the instability of QNMs. We use two different methods to illustrate the instability. First is the parametric flow, in which we confirm that the QNMs all migrate in unrestricted lines when the correction from $p$ or $q$ becomes remarkable. When $r_{+}$ is fixed, no matter $p$ or $q$, when they increase the average norms of QNMs will decrease. For parameter $m$, it hardly changes the imaginary part of QNMs while remarkably contributes to increase in the real part. This is quite different from previous studies on massive fields, in which $m$ is believed to reduce the energy leak and lead to quasi-resonance. Actually $m$ can even serve as a equivalent parameter as $l$. We then use the 2-norm of the matrix to define the pseudospectrum and reveal its characteristic through contour line plot, which show great accordance to the parametric migration flow. 

In the third part we study the evolution in the time domain both analytically and numerically. We first use the high-frequency approximation to obtain the expression of the two basic solution of the wave equation, and then the transfer coefficient and finally Green function. We illustrate how the contributions from different parts of contour integral influence the evolution. We then use numerical methods to clarify the impact caused by certain corrections. As expected, no power-law tail is found and $p$ shows more importance than $q$. Still, quantum correction can influence the oscillating rate and the lasting period of the Precursor stage. These are all potential observables in related experiments in the future. 

There still remains plenty of questions to answer:
\begin{enumerate}
    \item As $m$ has been found to have nearly opposite influence in QNMs from that of quantum correction and cosmological constant, could it be possible that the massive term $m$ precisely or approximately offsets the corrections caused by $p$ or $q$? If so, what is the specific parameter condition? This is very important in future detection if we want to determine the upper limit of parameters or oppositely, do parameter domain exclusion. 
   \item The detailed properties of the gap between a very small $p$ and $p=0$. In Fig.\ref{fig:qqq}, there are disturbing facts of immediate emerging new branches with rather small imaginary part when $p$ approaches 0, which means the characteristic quasi-resonance of massive fields in asymptotic flat spacetime are becoming remarkable gradually. Also, $p\rightarrow 0$ and $p=0$ majorly differs on whether there is branch cut and whether there is power-law tail, and this gap needs deeper understanding.
    \item How to attain the more precise approximation of the Green function as $|s|\rightarrow\infty$, as this determines how precisely we can predict the behavior in the Precursor stage.
    \item The strong cosmic censorship (SCC) hypothesis in qOS-dS spacetime has been tested under scalar perturbation in \cite{PhysRevD.109.064012}. The conclusion there is actually based on the conclusion of \cite{10.1063/1.4996575} on scalar perturbation in RNdS, which means that is just an approximation. If we want to repeat this in Proca perturbation, we had better know more details about the relation between QNMs of vector field and SCC in quantum corrected spacetime.
    \item The entire work is based on semi-classical field theory, but strict conclusions still need to be done on complete LQG framework. 
\end{enumerate}

\section{acknowledgements}
We thank Li-Ming Cao for guidance, Liang-Bi Wu, Yu-Sen Zhou and Long-Yue Li for comments and very necessary assists in the technical instructions, Anıl Zenginoğlu for his fundamental works in hyperbolodital framework.

\bibliography{reference}{}
\bibliographystyle{apsrev4-1}

\end{document}